\documentclass[preprint, 11pt,5p]{elsarticle}

\usepackage{amssymb}
\usepackage{verbatim}
\usepackage{amsmath}
\usepackage{multirow}
\usepackage{diagbox}
\usepackage{graphicx}
\newtheorem{prop}{Proposition}

\begin{document}

\begin{frontmatter}

\title{BAN-GZKP: Optimal Zero Knowledge Proof based Scheme for Wireless Body Area Networks\tnoteref{label0}}
\tnotetext[label0]{An extended abstract of this paper appeared in IEEE 14th International Conference on Mobile Ad Hoc and Sensor Systems (MASS), 2017}

\author[label1]{Gewu Bu}
\address[label1]{Sorbonne Universit\'e, UPMC, LIP6, CNRS UMR 7606}


\ead{gewu.bu@lip6.fr}

\author[label1]{Maria Potop-Butucaru}
\ead{maria.potop-butucaru@lip6.fr}


\begin{abstract}

BANZKP is the best to date Zero Knowledge Proof (ZKP) based secure lightweight and energy efficient authentication scheme designed for Wireless Area Network (WBAN). It is vulnerable to several security attacks such as the replay attack, Distributed Denial-of-Service (DDoS) attacks at sink and redundancy information crack. However, BANZKP needs an end-to-end authentication which is not compliant with the human body postural mobility. We propose a new scheme BAN-GZKP. Our scheme improves both the security and postural mobility resilience of BANZKP. Moreover, BAN-GZKP uses only a three-phase authentication which is optimal in the class of ZKP protocols. To fix the security vulnerabilities of BANZKP,  BAN-GZKP uses a novel random key allocation and a Hop-by-Hop authentication definition. We further prove the reliability of our scheme to various attacks including those to which BANZKP is vulnerable. Furthermore, via extensive simulations we prove that our scheme, BAN-GZKP, outperforms BANZKP in terms of reliability to human body postural mobility for various network parameters (end-to-end delay, number of packets exchanged in the network, number of transmissions). We compared both schemes using representative convergecast strategies with various transmission rates and human postural mobility. Finally, it is important to mention that BAN-GZKP has no additional cost compared to BANZKP in terms memory, computational complexity or energy consumption.

\end{abstract}


\end{frontmatter}


\section{Introduction}
\label{sec1}

Wireless Body Area Networks (WBAN) is a special  kind of Wireless Sensors Networks (WSN). In WBAN,  networked body sensors collect user's physiological data and transmit them to a sink node. There is a tremendous difference between WBAN and classical WSN. In WBAN nodes are distributed on/in human body and, similar to the Delay-Tolerance Networks (DTN),  move with the human postural mobility \cite{latre2011survey}, \cite{hayajneh2014survey}. Because of that, the network topology in Intra-WBAN dynamically changes following the postural body mobility. In a recent work related to channel modes for WBAN \cite{IEEEexample:naganawa2015simulation} the  authors advocate for the use of multi-hops communication in WBAN. 

Multi-hop WBAN communication schemes easily adapt to postural mobility  where nodes and links are highly dynamic \cite{IEEEexample:naganawa2015simulation}. Also, multi-hop communications need lower transmission power compared to one-hop direct communication where source nodes have to  use enough transmission power in order to make sure that their messages can reach the sink directly. Moreover, lower transmission powers automatically reduce the radio radiation of the human body, which became an  important issue today \cite{IEEEexample:naganawa2015simulation}.

However, multi-hop communication in WBAN is vulnerable to security and privacy attacks. Any medical data error, leakage or imitation may lead to a wrong medical treatment. The disclosure of critical health information can also have irreversible consequences on the patients daily life. Security mechanisms are thus  needed in WBAN to protect user's data from malicious eavesdropping, tampering or abu-
se. 

Recently,  the literature investigated Inter-WBANs security. As examples, \cite{ibrahim2016secure}, \cite{omala2017efficient} and \cite{bellekens2016pervasive} discuss the security mechanism for communications from the sink to the remote Health Centre (hospitals or online doctors). 
In this context the security of the Intra-WBAN communication should also be carefully considered: date leakage or tampering of source nodes from Intra-WBAN area leads to a meaningless subsequent Inter-WBAN security protection.

The challenges of Intra-WBAN security are threefold:
\begin{itemize}
\item The computing capacity of WBAN devices is limited. Traditional encryption and decryption algorithms used for personal computers or mobile phones may be not applicable as they are to the WBAN devices. 
\item Poor storage of WBAN devices may not be able to store too much shared content to make effective the recent complex authentication and security protocols. 
\item Control message exchanges may lead  to poor applicative performances.
\end{itemize}

\paragraph{Related Works}
The most basic encryption mechanism, symmetric encryption, uses the same secret key to encrypt and decrypt data. As symmetric key can be directly used in Stream cipher or Block cipher, the coding speed and its efficiency are very competitive. However, in symmetric encryption, by using the all-networks-widely fixed key, if one node has been compromised, the secret key will be known by adversary who can then monitor the entire networks. Also, symmetric encryption suffers from replay attack due to the use of the same encryption key. Some improvement solutions come out to solve the  replay attack problem. One example is MiniSec \cite{luk2007minisec}. Without using the same key, MiniSec uses data sequence as a part of encryption key. However, MiniSec needs to synchronize sequences of packets when the number of missing messages is important. This is often the case in a WBAN environment. Also MiniSec suffers from DDoS attack at the sink, since MiniSec doesn't force a hop-by-hop authentication, malicious message traffic from adversary can deliver to all the network. Other solutions \cite{li2013secure}, \cite{rushanan2014sok} and \cite{rehman2015ecc} come with specified \emph{Key Agreement Mechanism} to ensure the key will change periodically. However, most of works do not mention the networks performance impact when applied in a real WBAN environment.



Public Key Infrastructure (PKI) is a widely-used asymmetric encryption, authentication and access co-
ntrol mechanism (\cite{watro2004tinypk}, \cite{sahebi2016seecc}). Especially after the introduction of elliptic curve encryption (ECC) mechanism \cite{wang2011public}, which is proved more efficient than traditional PKI. However, this kind of mechanism needs an additional Certification Authority (CA) to generate user certification. ID-based (\cite{shamir1984identity}, \cite{qin2016security}) or certificateless based (\cite{al2003certificateless}, \cite{liu2014certificateless}) mechanisms need also the Networks Manager (NM) to achieve this security function, which are not well suitable for Intra-WBAN communication. Complex parameter assignment and key management are also the major challenges for asymmetric encryption in WBAN.

Another trend is the security scheme based on the physiological signal or channel quality introduced in \cite{ramakrishnan2015switch}, \cite{sampangi2012iamkeys}, \cite{choudhary2016robust} and \cite{peter2016design}. The nodes can use the collected physiological signals to encrypt and decrypt messages. However, the processing of these physiological signals needs additional powerful elements to handle. Because in this case, sensor nodes don't only need to store physiological signals as Data, but also need a further treating of these signals as a part of encryption process. These elements are expensive and consume additional energy. For example, in \cite{ramakrishnan2015switch}, to use Electrocardiography (ECG) signals, it needs additional device to do Discrete Meyer Wavelet transform (DWT). Also, the distance, the changing of temperature or the human body mobility can make the collected physiological signals different at two different nodes.

More recently, in order to respond to the three challenges of WBAN security, BANZKP \cite{chaudet2016banzkp} and Tin-
yZKP \cite{ma2014tinyzkp} where specifically designed for WBAN and use ZKP based authentication mechanism.

The best to date ZKP-based scheme, BANZKP \cite{chaudet2016banzkp},  uses less memory to store private secrets and requires less computing capacity than TinyZKP \cite{ma2014tinyzkp} and the Elliptic Curve Encryption Based Public Key Authentication scheme \cite{wang2011public}. BANZKP is also resilient to  a wide range of attacks. However, BANZKP still suffers from some specific malicious attacks such as Data Replay attack, DDoS Attack at sink and Redundancy Information Crack. Moreover, the resilience of BANZKP to human body postural mobility in WBAN environment was left as open question.

\paragraph{Our Contribution}
In this paper, based on an extensive analyze of the security weakness of BANZKP we propose a new ZKP-based scheme that outperforms BANZKP from both security and networking point of view. An extended abstract of this paper appear in \cite{bu2017ban}. Our scheme BAN-GZKP is resilient to Data Redundancy Cracking, Data Replay Attack and DDoS Attack at the sink and optimizes the ZKP exchanging scheme. Furthermore, we stress both BAN-GZKP and BANZKP schemes face to human body postural mobility. When these schemes are plugged to convergecast protocols we prove via extensive simulations that for strategies that use BAN-GZKP scheme outperforms with respect to the case when BANZKP is used. Additionally, our BAN-GZKP scheme presen-
ts better computational complexity and less energy consumption than BANZKP while it has the same memory complexity. 

\paragraph{Roadmap}
The paper is organised as follows. Section \ref{sectionI} presents and overview of the BANZKP scheme and discusses its vulnerabilities. Then, Section \ref{sectionII} presents our new BAN-GZKP, and its security analysis. The resilience of BANZKP and BAN-GZKP face to human body postural mobility when BANZKP and\\
BAN-GZKP are combined with known convergecast strategies, and theirs performances comparison analysis are shown in Section \ref{sectionIII}. We present, finally, a summary and conclude our work in the Section \ref{sectionIV}.

\section{BANZKP vs Security Attacks}
\label{sectionI}
In this section we recall briefly  BANZKP \cite{chaudet2016banzkp} (the best to date ZKP-based scheme designed for WBAN), then we analyze its vulnerabilities in terms of resilience to security attacks.
\subsection{BANZKP  Overview}
BANZKP \cite{chaudet2016banzkp} combines a  \emph{Zero knowledge Proof} and a \emph{Commitment Scheme}.

The \emph{Zero knowledge Proof} scheme ensures bidirectional authentication between two parties (a sender and a verifier). The idea is by exchanging challenge and response messages between two parties, they can finally trust each other that they hold the same \emph{secret information}, but none of them sent this secret information into the channel during the challenge and response phase. The security level is guaranteed by the fact that it is practically impossible to solve the discrete logarithms for numbers represented on hundreds of bits \cite{ma2014tinyzkp}. In BANZKP the two parties exchange \emph{five challenge/response} messages and never disclose the shared secret. 
   
 The \emph{Commitment Scheme} ensures that a sender transmits an encrypted message to a receiver who does not have the decryption key yet. The key is transmitted later as soon as the identity of the receiver is confirmed. In BANZKP, the Commitment Scheme is transmitted directly in plaintext. Because, this key is only used to verifier the identity of the sender and will be used only once per authentication session. So this key is useless after this session and will not give any secret information to anyone.

BANZKP  between two nodes $N_{1}$ and $N_{2}$ executes the following five steps:\\
\small{

$1)N_{1}>----E(K_{I}[ID_{N_{1}}||V^{p}])---->N_{2}$\\

$2)N_{1}<E(K_{I}[ID_{N_{2}}||V^{q}||RI]),E(K_{CS}[V_{RI}^{p*q}])<N_{2}$\\

$3)N_{1}>---E(K_{I}[ID_{N_{1}}||V_{RI}^{q * p}])---->N_{2}$\\

$4)N_{1}<------K_{CS}--------<N_{2}$\\

$5)N_{1}>---E(K_{I}[ID_{N_{1}}||DATA])--->N_{2}$\\

}

\normalsize
$ID_{N_{1}}$ and  $ID_{N_{2}}$  are identities of $N_{1}$ and $N_{2}$ respectively; $V$ is the shared secret number; $p$ and $q$ are two random values generated by $N_{1}$ and $N_{2}$, respectively; $K_{I}$ is a shared key between $N_{1}$ and $N_{2}$; $K_{CS}$ is a random key generated by $N_{2}$ for the \emph{Commitment Scheme} and the function $E(K[a])$ means encrypt $a$ with key $K$. $RI$ is the indicator of the beginning of an interval value of $V^{q * p}$, represented by $V_{RI}^{q * p}$. In BANZKP, the size of this interval is 200 bits.

Notice that, $V$ should be a number big enough, and $p$ and $q$ should also randomly chosen to be big enough. So that we can make sure the $V^{q * p}$ has minimal 1096 bits to randomly chose a interval of 200 bits \cite{chaudet2016banzkp}.

During the \emph{initialization} phase, both participant nodes store locally a shared secret number $V$ and a shared key $K_{I}$ by operator (user) manually. Considering the limited number of the nodes in WBAN, manual operations is feasible. 

\begin{figure}
\centering
\includegraphics[width=0.4\textwidth]{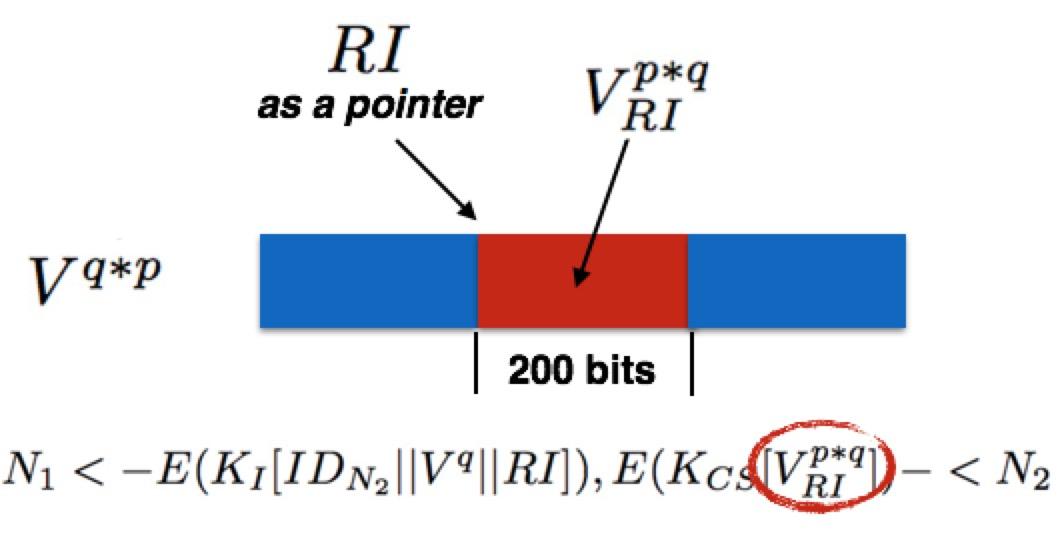}
\caption{Computing $V_{RI}^{p*q}$}
\label{RI}
\end{figure}

During the \emph{authentication} phase, when a source node has Data packet to send, an authentication session is initiated: 
\begin{enumerate}
\item $N_{1}$ initiates the authentication session. It choses a random value $p$ and computes $V^{p}$. It then encrypts his $ID$ and $V^{p}$ by $K_{I}$ and sends the whole message to $N_{2}$. 
\item Upon reception of $V^{p}$, $N_{2}$ generates a random value $q$ and computes $V^{q}$ and $V^{p*q}$. $N_{2}$ then generates a random indicator $RI$ and choses a 200 bits interval value of the $V^{p*q}$ from the indicator $RI$, as $V_{RI}^{p*q}$ (see Figure \ref{RI}). $N_{2}$ sends back to $N_{1}$: (1) $ID_{N_{2}}$, $V^{q}$ and $RI$ encrypted by $K_{I}$; (2) $V_{RI}^{p*q}$ encrypted by a random chosen session key $K_{CS}$.
\item Upon reception of response of $N_{2}$, $N_{1}$ computes $V^{q*p}$ and uses the received $RI$ to compute $V_{RI}^{q*p}$. $N_{1}$ then sends his $ID$ and $V_{RI}^{q*p}$ encrypted by $K_{I}$ to $N_{2}$. $N_{1}$ also keeps $E(K_{CS}[V_{RI}^{p*q}])$ from $N_{2}$ and waits the $K_{CS}$ sent later to verify the legitimacy of $N_{2}$. 
\item Upon reception of $V_{RI}^{q*p}$, $N_{2}$ compares this value with his own value, $V_{RI}^{p*q}$. If these two values are equal, then $N_{2}$ is sure that $N_{1}$ has the same shared secret $V$. Then it confirms the authentication by sending the $K_{CS}$ to $N_{1}$. Otherwise,  $N_{2}$ discards the message and closes the session.
\item Upon reception of $K_{CS}$, $N_{1}$ decrypts\\
$E(K_{CS}[V_{RI}^{p*q}])$ and compares this value with its own value, $V_{RI}^{q*p}$. If these two values are equal, $N_{1}$ is sure that $N_{2}$ has the same secret $V$, and sends $ID$ and $DATA$ encrypted by $K_{I}$ to $N_{2}$. Otherwise, $N_{1}$ discards the message and closes the session.
\end{enumerate}

 BANZKP copes with the following attacks:
\begin{itemize}
\item{\emph{Forge Nodes} \cite{chaudet2016banzkp}: } Thanks to the bidirectional authentication, any forge node attempting to disguise itself in a legitimate node cannot be certified. This is due to the fact that forge node has no information on the shared secret. Hence, it cannot compute the correct authentication response. 

\item{\emph{Replay Attack} \cite{chaudet2016banzkp}: } Adversary could intercept previous exchanged messages and try to use them to make other nodes in the networks trust its identity and finish the  bidirectional authentication. The use of randomly chosen $p$ and $q$ makes each authentication session different with respect to the previous ones. Hence, old messages cannot help to correctly execute the authentication.

\item{\emph{Man in the Middle Attack} \cite{chaudet2016banzkp}: } In this attack, the adversary listens channels and try to steal the shared secret. BANZKP does not send directly secret information. 

\item{\emph{Guessing Attack} \cite{chaudet2016banzkp}: } The use of random values for $q$, $p$ and $RI$ makes practically impossible  for the adversary to guess the shared secret value $V$ from $V_{RI}^{q*p}$ or $V_{RI}^{p*q}$.

\item{\emph{Privacy Attack} \cite{chaudet2016banzkp}: } The adversary may try to eavesdrop. BANZKP prevents this attack by encrypting exchanged Data message with $K_{I}$.
\end{itemize} 

\subsection{BANZKP vulnerabilities}
In this section we analyze the BANZKP vulnerabilities. 
\bigskip

\emph{Data Replay Attack: }
BANZKP scheme can prevent malicious authentication message replay by using the random values $q$, $p$ and $RI$. However, for encrypting Data message, a constant key $K_{I}$ is used for all Data message. A conscious adversary may launch a Data Replay Attack by observing the pattern of the exchanges. For example, two nodes are exchanging the authentication messages; an adversary, who holds a captured previous Data message encrypted by $K_{I}$ from $N_{1}$ in previous authentication session between $N_{1}$ and $N_{2}$, is listening the channel. In the phase 4) of BANZKP, $N_{2}$ sends the random key, $K_{CS}$ to $N_{1}$. The adversary can also receive this key.  At this particular moment, the adversary knows that $N_{2}$ is, from now on, waiting for an encrypted Data. The adversary thus sends immediately the previous captured Data message to $N_{2}$ to pretend this expired Data message as a fresh one. The consequence is that $N_{2}$ treats the expired Data message as the right one and ignores the right message from $N_{1}$ and allows the adversary to inject invalid Data into the network.
\bigskip

\emph{Redundancy Information Crack: }
The encryption in BANZKP uses the stream cipher mechanism where each bit of collected Data does the exclusive or with each bit of the encryption key. Since the key used for Data encryption is always the same $K_{I}$ at the end of each authentication session, Data messages sent by source nodes have the following format: $M_{1} = Data_{1}$ xor $K_{I}$, $M_{2} = Data_{2}$ xor $K_{I}$... By capturing $M_{1}$ and $M_{2}$, the adversary can do the $xor$ of them to get redundancy information: $M_{1}$ xor $M_{2}$ $=Data_{1}$ xor $Data_{2}$. After getting enough redundancy information, encrypted Data could be cracked and from the Data, $K_{I}$ then will be no longer a secret.
\bigskip

\emph{DDoS Attack at Sink: }
 BANZKP was designed to work for both single-hop and multi-hop WBAN networks. 
In multi-hop WBAN environment, BANZKP uses relay nodes to forward the source messages to the sink. From the original BANZKP design, the bidirectional authentication is an end-to-end authentication between the source node and the sink. Relay nodes will just forward  the messages. Hence all the authentication or Data information is transparent to them. If an adversary sends continuous invalid authentication request messages (phase 1 of the BANZKP scheme), relay nodes will forward these messages to the sink. The sink will then suffer from a DDoS attack if the amount of the authentication requests is high. The network resources will be consumed by these invalid authentication requests and the real authentication messages get thus less chance to reach the sink.
\bigskip

\emph{Potential Adaptation Problem: }
Intra-WBAN co-
mmunication is affected by high nodes mobility, the important channel attenuation given by the signal absorption and the reflection of human body. An end-to-end authentication may not be efficient when facing the unstable and high dynamic environments due to packets loss during the multi-hops transmission from sources to the sink. Any loss of timeout during the transmission will lead to the fail of the whole authentication phase.

\section{BAN-GZKP}
\label{sectionII}
Original BANZKP scheme shows vulnerabilities in terms of security and reliable communications. In this section we present our new BAN-GZKP scheme that improves over BANZKP in several ways. BAN-GZKP is resilient to all the attacks supported by BANZKP plus the Data Replay Attack, Redundancy Information Crack and DDoS attack at sink. Moreover,  BAN-GZKP presents better performances in terms of percentage of packets received at the sink, end-to-end delay and the number of transmissions. BAN-GZKP needs only \emph{three phase exchanges} which is optimal in the ZKP class of schemes.
 
We first present the ingredients that compose our new BAN-GZKP scheme and analyze its resilience attack and its complexity in terms of memory and computation. 
\subsection{BAN-GZKP Ingredients}
In order to tolerate Data Replay Attack, Redundancy Information Crack and DDoS attack at sink BAN-GZKP uses three ingredients: \emph{a random key allocation}, a \emph{hop-by-hop} authentication and the ZKP \emph{Exchanging Schemes Optimization}.
\subsubsection{Random Key Allocation}
Data Replay Attack and Redundancy Information Crack are possible in BANZKP because a constant key $K_{I}$ is used to encrypt all Data messages. 

An effective and well adapted key management mechanism is necessary to generate different encryption keys for Data messages per session.  

\begin{figure}
\centering
\includegraphics[width=0.4\textwidth]{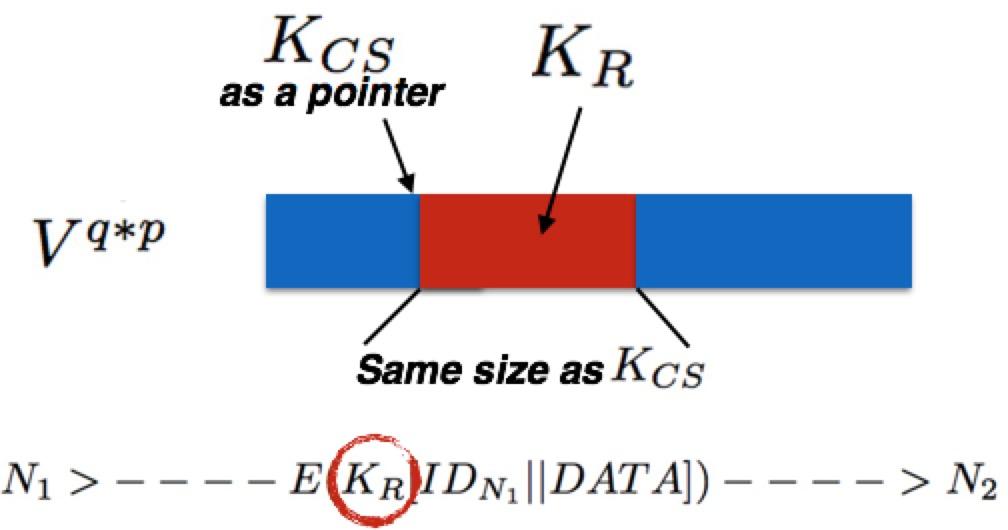}
\caption{Computing $K_{R}$, a Random Data encryption Key}
\label{KR}
\end{figure}

The idea of Random Key Allocation is as follows: when nodes authenticate, the shared secret value $V^{q*p}$ will be obligatory computed for each authentication session. Since $p$ and $q$ are randomly chosen, $V^{q*p}$ is also random. During the authentication Phase 4 in the original BANZKP, $N_{2}$ will send the random session key to $N_{1}$ to decrypt previous information. Notice that, even though $K_{CS}$ is random, this key should not be used  to encrypt Data messages because it has been sent on clear text. Our idea is to use $K_{CS}$ as  a random pointer that will point to a bit in the binary representation of the random value $V^{q*p}$. Then we chose an interval in the binary representation of $V^{q*p}$ that starts with the bit pointed by the random pointer $K_{CS}$. This interval, of length $K_{CS}$  can be seen as a random key, $K_{R}$, to encrypt Data message for the current session (see Figure \ref{KR}).

Our Random Key Allocation does not require additional keys at the initialization and does not need the send of additional fields.

\subsubsection{Hop-by-Hop Scheme}
Note that Sink-Side DDoS Attack happens in the end-to-end authentication scheme because relay nodes cannot detect whether the authentication message is legal or not. Only the sink can do. To solve this problem and prevent Sink-Side DDoS Attack, we need to provide relay nodes with the capacity to detect invalid authentications.

The idea is as follows, instead of doing the authentication between the pair source-sink, we let source nodes to initiate authentication directly with their one-hop  neighbours. After this authentication phase finishes with success, a source is allowed to send Data messages to the authenticated neighbour. The neighbour who receives Data messages can then initiate authentication with its neighbours until Data reaches to the sink.

An adversary who wants to initiate a large number of invalid authentication requests to block the network will be detected directly by its one-hop neighbours and the DDoS Attack can thus be limited in a local range.

\subsubsection{ZKP Exchanging Scheme Optimization}
%

Original ZKP schemes need five-times continuously message exchanging to achieve a bidirectional authentication. This scheme could be optimised to three-times continuously message exchanging under certain conditions. The reduction of the exchanged messages could save network resources, and further improve the total Intra-WBAN performance. The idea of the optimization proposed for BAN-GZKP is as follows:

\emph{A)} For any authentication exchanging between two nodes who never be authenticated to each other before, we take exactly the same scheme as the original BANZKP to do the bidirectional authentication.

\emph{B)} When a source node $N_{1}$ initiates authentication with another node $N_{2}$ that previously authenticated with $N_{1}$ and that recognizes the identity of $N_{1}$, we can then optimise the total authentication scheme to the following:\\
\small{

$1)N_{1}>-----E(K_{I}[ID_{N_{1}}||V^{p}])----->N_{2}$\\

$2)N_{1}<--E(K_{I}[ID_{N_{2}}||V^{q}||RI||R||V_{RI}^{p*q}])--<N_{2}$\\

$3)N_{1}>----E(K_{R}[ID_{N_{1}}||DATA])---->N_{2}$\\ 

}

\normalsize
When a source node has Data packet to send, an authentication session is initiated:
\begin{enumerate}
\item $N_{1}$ initiates the authentication session. It choses a random value $p$ and computes $V^{p}$. It then encrypts its $ID$ and $V^{p}$ by $K_{I}$ and sends the whole message to $N_{2}$. 
\item $N_{2}$ recognizes the identity of $N_{1}$, then  $N_{2}$  instead of sending back $E(K_{I}[ID_{N_{2}}||V^{q}||RI]), \\E(K_{CS}[V_{RI}^{p*q}])$, where $V_{RI}^{p*q}$ is encrypted with $K_{CS}$ (as in original BANZKP scheme), it sends back directly $V_{RI}^{p*q}$ encrypted with the initial key $K_{I}$.  In our BAN-GZKP $N_{2}$ needs just to send  a random pointer $R$ for the Random Key Allocation. Hence, the final message sent back to $N_{1}$ is: $E(K_{I}[ID_{N_{2}}||V^{q}||RI||R||V_{RI}^{p*q}])$.
\item After receiving the response of $N_{2}$, $N_{1}$ finishes the authentication using the same mechanism, and choses a random key, $K_{R}$, from the pointer $R$ of Random Key Allocation and encrypt Data by $K_{R}$ then sends the message to $N_{2}$, if $V_{RI}^{p*q}$ and $V^{q*p}$ are equal. We thus can complete the authentication session after the first successful authentication between these two nodes. If not, $N_{1}$ discards the message and closes the session.
\end{enumerate}

\subsection{BAN-GZKP Security Analysis}
BAN-GZKP reduces the number of authentication messages and also tolerates the attacks tolerated  by BANZKP scheme. Additionally it tolerates Data Replay Attack and Redundancy Information Crack. As for the case of BANZKP, BAN-GZKP can be implemented either end-to-end or hop-by-hop. Note that BAN-GZKP hop-by-hop scheme is also resilient to Sink-Side DDoS Attack.

Inspired by \cite{dolev1983security}, we propose formal security proof of BAN-GZKP. We first define a node $A$ hold $S$ as node $A$ known information is:
\begin{align} 
	I_{A} = \{S\}
\end{align}
Let the operation of sending a message including $X$ as content of the message from node $A$ to node $B$ at $i$th authentication phase is: 
\begin{align} 
	A:(A, X, B)_{i}, \quad i \in \overline{1,5}
\end{align}
In BAN-GZKP, for the first time of the authentication, we need total five authentication phases (each authentication message exchange is seen as a authentication phase), but form the second time of the authentication between these two nodes, it's need only three authentication phases.
Let the Checking operation of a legitimate (honest) node $A$ to verifier a received message from $B$ is legal or not is:
\begin{align} 
	Checking_{A}((B, X, A)_{i})
\end{align}
If message $(B, X, A)_{i}$ can not be decryption correctly or $X$ can not is not the correct response of previous challenge, then the Checking operation failed, node drop the received message. If not, Checking succeeded, then node continues the next authentication exchange.

Let's define a node $Z$ as a smart adversary, who apply operations as following:
\begin{itemize}
\item{$Z$ can intercept the message sent form $A$ to $B$ at any authentication phase.}
\item{$Z$ can initiate an authentication message: $Z:(A, X, B)_{i}$ where $A$ and $B$ can be any nodes in the network, and X can be anything belonging to $I_{Z}$.}
\item{$Z$ can don't check any message and confirm arbitrarily that checking failed or succeeded.}
\item{$Z$ can update $I_{Z}$, when intercepting message.}
\item{$Z$ can try to decrypt encrypted message by using information from $I_{Z}$ at any time.}
\end{itemize}

In the following, we prove that BAN-GZKP is tolerant to the adversary $Z$ defined above.

\begin{prop}
\emph{BAN-GZKP is resistant to Forge}\\
\emph{Nodes Attack.}
\end{prop}
\textbf{Proof} Let $Z$ is a Forge Nodes attempting to disguise itself in a legitimate node to authenticate with $B$. The initial information of $Z$ and $B$ is:\\
$I_{Z}$ = $\{ID_{Z}, ID_{B}, K_{I_{Z}}, V_{Z}, p\}$\\
$I_{B} = \{ID_{B}, K_{I_{B}}, V_{B}\}$\\
And following the operations shown below, $Z$ can not authentication succeed with $B$.\\
$1)$ $Z:(Z, E(K_{I_{Z}}[ID_{Z}||V_{Z}^{p}]), B)_{1}$\\
$2)$ $Checking_{B}((Z, E(K_{I_{Z}}[ID_{Z}||V_{Z}^{p}]), B)_{1})$\\
$3)$ $Checking$ $Failed:$ $K_{I_{Z}}$ $\neq$ $K_{I_{B}}$\\
As $Z$ has only information about the ID of $B$ and itself, Z can not pretend to be a legitimate node, because the message encryption key $K_{I_{Z}}$ is chosen arbitrarily, $B$ can not decrypted correctly, the checking will fail. As the sender node will always do the authentication with the receiver node; the receiver node does also the authentication with sender node only the first time. In this case, even though the adversary can disguise itself in a legal node $A$  who previously finished the authentication with the receiver $B$. Because this message sent by this adversary cannot be decrypted correctly.

\begin{prop}
\emph{BAN-GZKP is resistant to Authentication Replay Attack.}
\end{prop}
\textbf{Proof} The proof decomposes in two parts. In the first parts, let $Z$ intercept a message $(A, X, B)_{1}$. Then $Z$ can replay directly this message to disguise itself in a the node $A$ who previously finished the authentication with $B$. The initial information of $Z$ and $B$ are:\\
$I_{Z}$ = $\{ID_{Z}, ID_{A}, ID_{B}, (A, X, B)_{1}, K_{R_{Z}}\}$\\
$I_{B}$ = $\{ID_{A}, ID_{B}, K_{I_{A}} = K_{I_{B}}, V_{A} = V_{B}, K_{R_{B}}\}$\\
Following the operations shown below, $B$ can detected the message is a replay message.\\
$1)$ $Z:(A, X, B)_{1}$\\
$2)$ $Checking_{B}((A, X, B)_{1})$\\
$3)$ $Checking$ $Succeeded:$ $K_{I_{A}} = K_{I_{B}}, V_{A} = V_{B}$\\
$4)$ $B:(B, E(K_{I_{B}}[ID_{B}||V_{B}^{q}||RI||R||V_{RI}^{p*q}]), A)_{2}$\\
$5)$ $Z$ $intercepts$ $(B, E(K_{I_{B}}[ID_{B}||V_{B}^{q}||RI||R||V_{RI}^{p*q}]), A)_{2}$\\
$6)$ $Z$ $confirm$ $Checking$ $Succeeded$\\
$7)$ $Z: (A, E(K_{R_{Z}}[ID_{A}||DATA]), B)_{3}$\\
$8)$ $Checking_{B}((A, E(K_{R_{Z}}[ID_{A}||DATA]), B)_{3})$\\
$9)$ $Checking$ $Failed:$ $K_{R_{Z}}$ $\neq$ $K_{R_{B}}$\\
Even though $B$ trusts the replay message, the Data message encrypted with the wrong encryption key $K_{R_{Z}}$ will  be detected by $B$ at the end of the authentication phase.

In the second part, we assume that the adversary tries to replay the message sent by $B$ in phase 2, and try to get information from $A$, the initial information of $Z$ and $A$ are:\\
$I_{Z}$ = $\{ID_{Z}, ID_{A}, ID_{B}, (B, X, A)_{2}\}$\\
$I_{A}$ = $\{ID_{A}, ID_{B}, K_{I_{A}} = K_{I_{B}}, V_{A} = V_{B}\}$\\
Following the operations shown below, this message will directly drop since $p$ chosen by $A$ are different random values for each session, a replay message can not pass the checking at $A$.\\
$1)$ $A:(A, E(K_{I_{A}}[ID_{A}||V_{A}^{p}]), B)_{1}$\\
$2)$ $Z$ $intercepts$ $(A, E(K_{I_{A}}[ID_{A}||V_{A}^{p}]), B)_{1}$\\
$1)$ $Z:(B, X, A)_{2}$\\
$2)$ $Checking_{A}((B, X, A)_{2})$\\
$3)$ $Checking$ $Failed:$ $V_{RI}^{p*q}$ $computed$ $from$ $A$ $is$ $different$ $from$ $V_{RI}^{p*q}$ $received$ $from$ $X$\\
Notice that, we cannot simplify the authentication at $A$. Otherwise, the adversary can replay the same message from $B$, to force $A$ to use the same key $K_{R}$ to encrypt the Data message. Hence, the adversary can later initiate a Redundancy Information Crack.

\begin{prop}
\emph{BAN-GZKP is resistant to Man in the Middle Attack and Guessing Attack.}
\end{prop}
\textbf{Proof} $Z$ can intercept messages exchanged between $A$ and $B$ as a Man in the Middle to try Guessing secret value $V$ hold in $A$ and $B$, The initial information of $Z$ is:
$I_{Z}$ = $\{ID_{Z}, ID_{A}, ID_{B}\}$\\
Following the operations shown below, $Z$ can not get any useful information to decrypt message and Guess the content of Data message.\\
$1)$ $A:(A, X_{1}, B)_{1}$\\
$2)$ $B:(B, X_{2}, A)_{2}$\\
$3)$ $A:(A, X_{3}, B)_{3}$\\
$...$\\
$x)$ $Z$ $update$ $I_{Z}$\\
$x+1)$ $I_{Z}$ = $\{ID_{Z}, ID_{A}, ID_{B}, X_{1}, X_{2}, X_{3}...\}$\\
$x+2)$ $Z$ $try$ $decrypt$ $messages$ $to$ $get$ $V$\\
The attempting of decryption will fail because none of information in $I_{Z}$ = $\{ID_{Z}, ID_{A}, ID_{B}, X_{1}, X_{2}, X_{3}...\}$ contain the secret number $V$.

\begin{prop}
\emph{BAN-GZKP is resistant to Data Replay Attack.}
\end{prop}
\textbf{Proof} $Z$ can replay a Data messages from $A$ and $B$, and make $B$ think the replay message is correctly message. The initial information of $Z$ and $B$ are:\\
$I_{Z}$ = $\{ID_{Z}, ID_{A}, ID_{B}, (A, X, B)_{3}\}$\\
$I_{A}$ = $\{ID_{A}, ID_{B}, K_{I_{A}} = K_{I_{B}}, V_{A} = V_{B}\}$\\
Following the operations shown below, the replay message will directly drop. Since in each session, $K_{R}$ is different, message encrypted by $K_{R}$ computed from other session can not be decrypted correctly by $B$:\\
$1)$ $Z:(A, X, B)_{3}$\\
$2)$ $Checking_{B}((A, X, B)_{3})$\\
$3)$ $Checking$ $Failed:$ $K_{R}$ $computed$ $from$ $B$ $can$ $not$ $decrypt$ $(A, X, B)_{3}$\\

\begin{prop}
\emph{BAN-GZKP is resistant to Redundancy Information Crack.}
\end{prop}
\textbf{Proof} $Z$ can intercept encrypted Data messages sent from $A$ and $B$, and make try to decrypt Data message. The initial information of $Z$ is:\\
$I_{Z}$ = $\{ID_{Z}, ID_{A}, ID_{B}\}$\\
Following the operations shown below, $Z$ can not get Data information from collecting Redundancy Information of encrypted Data messages.\\
$1)$ $A:(A, X_{1}, B)_{3}$\\
$2)$ $A:(A, X_{2}, B)_{3}$\\
$3)$ $A:(A, X_{3}, B)_{3}$\\
$...$\\
$x)$ $Z$ $update$ $I_{Z}$\\
$x+1)$ $I_{Z}$ = $\{ID_{Z}, ID_{A}, ID_{B}, X_{1}, X_{2}, X_{3}...\}$\\
$x+2)$ $Z$ $try$ $decrypt$ $messages$ $to$ $get$ $V$\\
As $X_{i}$ is the Data information encrypted by $K_{R}$ and in each authentication session, $K_{R}$ change randomly. $Z$ can not get Data information from Redundancy Information of encrypted Data message.

\begin{prop}
\emph{BAN-GZKP is resistant to DDoS attack at sink.}
\end{prop}
\textbf{Proof} a $Z$ may continue send message into its neighbour node to lance a DDoS attack at the sink by let neighbour node forwarding these message to the sink. The initial information of $Z$ and $B$ is:\\
$I_{Z}$ = $\{ID_{Z}, ID_{B}\}$\\
Following the operations shown below, all the useless transmission initialed from $Z$ will be blocked at $B$.\\
$1)$ $Z:(Z, X, B)_{1}$\\
$2)$ $Checking_{B}((Z, X, B)_{1})$\\
$3)$ $Checking$ $Failed:$ $B$ $can$ $not$ $decrypt$ $(Z, X, B)_{1})$\\
As $(Z, X, B)_{1}$ failed the checking at $B$, this message will be dropped directly. $B$ thus prevent the DDoS broadcasting into the whole network.

\subsection{Memory and Computational Complexity and Energy Consumption}
In  \cite{chaudet2016banzkp}, authors prove that BANZKP  improves over existing similar schemes in terms of memory requirements, computation complexity and energy consumption.
In the following we study the costs of BAN-GZKP compared to BANZKP.
In terms of the parameters required to be stored by each node for the initial phase, both the end-to-end and hop-by-hop BAN-GZKP  need  that source nodes and the sink store the shared value $V$ and the initial key $K_{I}$. 
Hence, BAN-GZKP has the same memory complexity as the original BANZKP.

In terms of \emph{computational complexity}, a complete authentication phase in the original BANZKP require-
s four times big number multiplications and five times encryption/decryption. Our BAN-GZKP scheme requires four times big number multiplications, but only three times encryption/decryption. Our scheme pres-
ents a better computation complexity for each complete authentication phase.

In terms of \emph{energy consumption}, the original\\ BANZKP needs five  transmission phases for a complete authentication. Even though our optimal scheme sends an additional field, $R$ as a random pointer in exchange phase number 2), BAN-GZKP needs only three transmission phases instead of five. The energy needed to send the $R$ value is hence negligible compared to two complete transmissions of BANZKP. 

To sum up, the new BAN-GZKP scheme optimizes BANZKP by adding a Random Key Allocation mechanisms and a Hop-By-Hop authentication. Moreover BAN-GZKP has better computational complexity and energy consumption than BANZKP.

\section{Analysis of Resilience to Postural Mobility}
\label{sectionIII}
In the following we analyze the effectiveness of BANZKP and BAN-GZKP schemes face to postural mobility. We therefore consider as case study the convergecast problem where Data messages sent by  source nodes are collected by a specific node in the network called sink. We enrich representative convergecast strategies specifically designed for multi-hop WBAN mentioned in \cite{badreddine2017convergecast} and \cite{bu2017total} with BANZKP and BAN-GZKP schemes, respectively. Note that the original BANZKP scheme requires an end-to-end authentication where all the authentication messages are transparent to relay nodes. Only the source and the sink can understand these messages;
BAN-GZKP is a Hop-By-Hop authentication scheme, where source nodes initiate authentication with their one hop neighbours. If these nodes are chosen to relay Data messages  then  before relaying these messages they apply the hop-by-hop authentication with their neighbours.

We evaluate the performances of both BANZKP and BAN-GZKP when these schemes are used in a secure convergecast process. Our evaluation focuses the percentage of packets received at sink for various rates of transmissions and various postural mobilities. 

In the next section we briefly present the convergecast strategies we evaluate and the way we plugg-
ed the BANZKP and BAN-GZKP schemes to these strategies. Then we discuss our simulation results.

\subsection{Convergecast Strategies} 
In \cite{badreddine2017convergecast} and \cite{bu2017total}, authors classify existing convergecast strategies for WBAN into five classes: Multi-Paths based Strategies, Tree-based Strategy, Dynamic Path Strategies, Gossip-based Strategies and\\
Attenuation-based Strategies. In our study we plug the BANZKP scheme on five different convergecast strategies (one representative per class).
\begin{itemize}
\item{\emph{Multi-Paths based Strategies: }}  are based on pre-determined paths and use these overlay paths as a reliability mechanism. An example is \emph{All Parents to All Parents} (APAP) strategy \cite{badreddine2017convergecast}. In APAP, each source node sends a message to maximum two pre-determined parents. Each parent then forwards received messages to maximum two of their parents.

\item{\emph{Tree-based Strategy: }} \cite{bu2017total} pre-constructs seven Best-Path Trees for different human postures shown in Figure \ref{walk1}.  Source nodes send messages through these paths to the sink. The pre-constructed Best-Path Trees are computed according to random attenuation distribution of each links. 

\item{\emph{Dynamic Path Strategies: }}  construct and update a tree-based overlay. The Collection Tree Protocol (CTP) \cite{colesanti2011collection} is an example of this class. In CTP each node sends additional BEACON messages to update the overlay route from each source to the sink.

\item{\emph{Gossip-based Strategies: }} use flooding. In this class we choose FloodToSink \cite{badreddine2017convergecast}, where a source diffuses messages to all its neighbours, then continue to  forward messages to all their neighbours and so on. In this case, every packet has a parameter, Time to Life (TTL), to limit the number of forwarding.

\item{\emph{Attenuation-based Strategies: }} these strategies are based on the negotiation of the channel attenuation. When a source has packets to send, it broadcasts first a Request (REQ) to ask an estimate attenuation from the receiver to the sink. The receiver of the Request will then send back a Reply (REP) with the required estimate attenuation value. The source will chose the next hop among replying nodes and sends data packets to the chosen one. In this class we investigate strategy MiniAtt \cite{badreddine2017convergecast}. This strategy choses one node who has the minimal estimate attenuation to the sink; if no Reply has been received for a while, the source will re-send the Request.
\end{itemize}

In our simulations we use five strategies to represent each class of convergecast strategies: APAP for Multi-Paths based Strategies; FloodToSink for Gossip-based Strategies; MiniAtt for Attenuation-ba-
sed Strategies; CTP for Dynamic Path Strategies and Tree-based Strategy to represent itself.

\subsection{How BANZKP and BAN-GZKP plugg to convergecast strategies}
We explain in Section \ref{BANzkp} and \ref{BANgzkp}, how BANZ-
KP and BAN-GZKP schemes can be plugged to the WBAN convergecast strategies, respectively. The general idea is that before each source node sends any Data packet, it needs to do the authentication with the sink in the BANZKP or with the next hop in the BAN-GZKP, respectively. 

\subsubsection{Convergecast with BANZKP}
\label{BANzkp}
In the authentication phase of BANZKP, source nodes need to exchange an authentication message with the sink. However as original convergecase strategies care about only how to flow authentication messages and Data messages from source to the sink (up stream), we need  to define how messages flow from the sink down to the source (down stream). In APAP, CTP  and Tree-based strategies, messages generated by the sink will follow the opposite route with respect to the up stream exchanges. That is, parents forward messages to their sons until messages reach the  sources. 

For MiniAtt strategy, for both up stream and down stream, nodes always need  from their neighbours attenuation information in order to chose the next hop. The difference is that for up stream, nodes ask the attenuation between the receiver and the sink; for down stream, nodes ask the attenuation information between the receiver and the initial source.

For FloodToSink there is no difference between the up stream and the down stream.

\subsubsection{Convergecast with BAN-GZKP}
\label{BANgzkp}
In BAN-GZKP, there is only up stream for Data from the source to the sink, since nodes only authenticate with their one hop neighbours. After the authentication, nodes send Data messages to the authenticated neighbour. So there is no authentication flow during the transmission, only the Data message will be forwarded from the source to the sink as up stream. 

Note that, for APAP and FloodToSink, there is always a multi-receiver when a source initiates the authentication. Receivers will reply to the source, the source then continue the authentication exchanging with all of them. But only the neighbour who finished the authentication phase firstly can be chosen as the legal next hop to avoid additional Data message and to respect the original convergecast strategy.

For MiniAtt strategy, the BAN-GZKP scheme can be integrated into the original Attenuation Require-Response scheme as follows: after a source chooses the next hop it begins to initiate the authentication directly with the chosen node.

For CTP and Tree-based strategies, there is only one parent to forward messages. Hence, the authentication is initiated by nodes with their parents. 

\subsection{Channel and Human Mobility Model}
The WBAN model we used in our research is proposed in \cite{badreddine2015broadcast}. 
They implement the realistic channel model proposed in \cite{IEEEexample:naganawa2015simulation} over the physical layer implementation provided by the Mixim framework \cite{kopke2008simulating}, who provides several extensions of existing simulation frameworks specified for wireless and mobility simulation. This channel model of an on-body $2.45\,GHz$ channel between $7$ nodes, that belong to the same WBAN, uses small directional antennas modeled as if they were $1.5cm$ away from the body. Nodes are assumed to be attached to the human body on the head, chest, upper arm, wrist, navel thigh, and ankle. In the convergecast strategies we consider six source nodes to send Data as follows: 0) navel,  2) head, 3) upper arm, 4) ankle, 5) thigh and 6) wrist, and one sink node that collects Data, node 1) chest.
 
\begin{figure}
\centering
\includegraphics[width=0.5\textwidth]{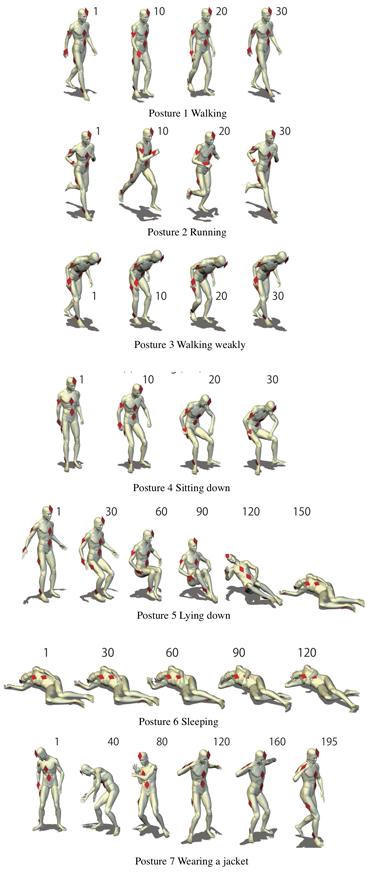}
\caption{7 Different Human Postures \cite{IEEEexample:naganawa2015simulation}}
\label{walk1}
\end{figure}

Nodes positions are calculated in $7$ postures: walking (walk), running (run), walking weakly (weak), sitting down (sit), wearing a jacket (wear), sleeping (sleep), and lying down (lie). Figure \ref{walk1} shows the positions of sensor nodes change with human mobility human postures within a time period in different postures model \cite{IEEEexample:naganawa2015simulation}. In each posture, a continuous human action has been device into many frames. Each single human body picture with a corresponding frame number, $x$, in a posture is a screenshot of this continuous human action at the $x$th frame. For example, in posture 1 Walking, the continuous action takes 30 frame, and in the Figure \ref{walk1}, it shows four screenshots at 1st frame, 10th frame, 20th frame and the 30th frame, respectively. The red diamonds in the figures represent sensors on the human body moving with the human mobility.

Channel attenuations are calculated between each couple of nodes for each of these positions as the average attenuation (in dB) and the standard deviation (in dBm). The model takes into account: the shadowing, reflection, diffraction, and scattering by body parts.
Naganawa et al. \cite{IEEEexample:naganawa2015simulation} studied a cooperative transmission scheme: two-hop relaying scheme. Using the simulated path loss, the performance of such scheme were evaluated by comparing the outage probability using different relay nodes against a direct link between a source and a destination. They advocate for the use of multi-hops communication which has the additional feature that it significantly decreases the transmission power. We thus interested in the network performance when applying BANZKP and BAN-GZKP in different multi-hops on-human communications.

\subsection{Simulation Results}
In order to evaluate the strategies described above in a realistic WBAN scenario, we implemented them under the OMNeT++ simulator enriched with  the Mixim project~\cite{kopke2008simulating} that specifically models the lower network WBAN layers.

We use standard IEEE 802.15.4 protocol as MAC layer. Note that the most recent standard\\
IEEE 802.15.6 proposed for WBAN  considers a star network topology (one hop) and does not  take into account the human body postural mobility. As stressed in the introduction we focus multi-hop networks and human body postural mobility.

We consider the following packet rates at the application layer: 1 packet/second, 5 packets/second and 10 packets/second. These values are commonly used in WBAN \cite{khan2010wireless}. The sensibility of WBAN devices is -100dBm and the transmission power has been set to -60dBm. We stress the studied strategies under a realistic channel model and postural mobility as described above.

\begin{figure}
\centering
\includegraphics[width=0.4\textwidth]{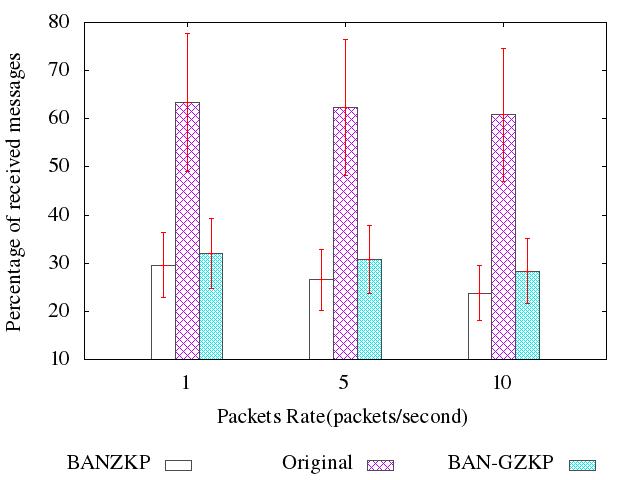}
\caption{Percentage of received messages for APAP in posture  1}
\label{Rp1APAP}
\end{figure}

\begin{figure}
\centering
\includegraphics[width=0.4\textwidth]{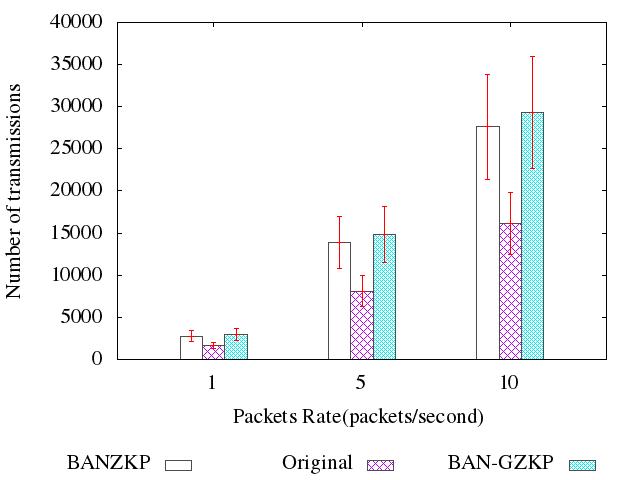}
\caption{Number of transmissions for APAP in posture 6}
\label{Tp6APAP}
\end{figure}

\begin{figure}
\centering
\includegraphics[width=0.4\textwidth]{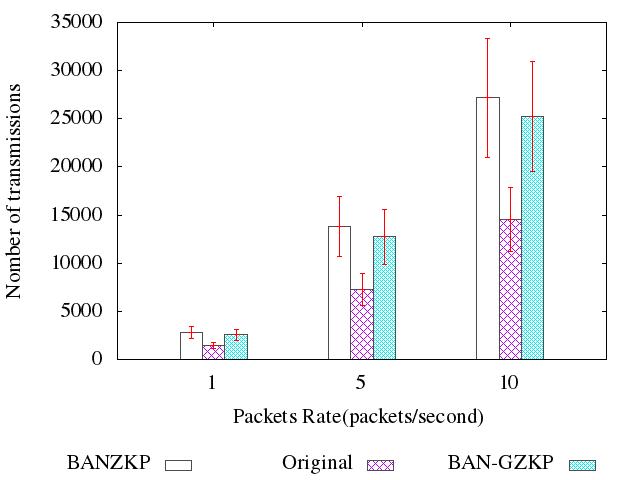}
\caption{Number of transmissions for APAP in posture 7}
\label{Tp7APAP}
\end{figure}

\begin{figure}
\centering
\includegraphics[width=0.4\textwidth]{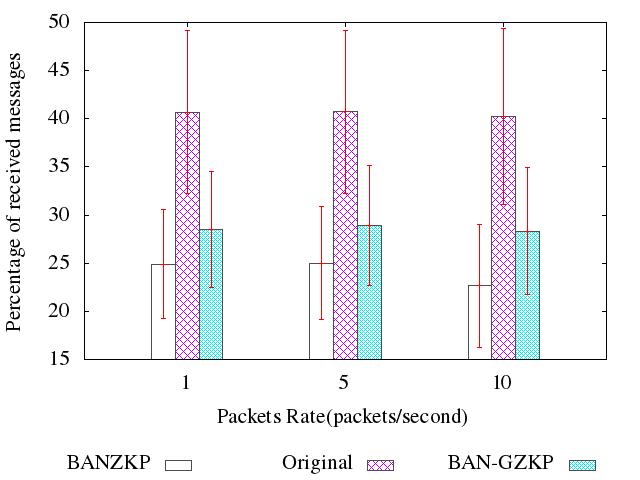}
\caption{Percentage of received messages for CTP in posture  1}
\label{Rp1CTP}
\end{figure}

\begin{figure}
\centering
\includegraphics[width=0.4\textwidth]{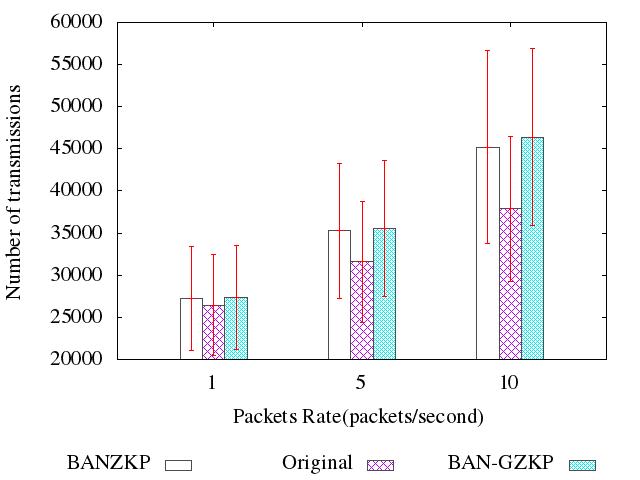}
\caption{Number of transmissions for CTP in posture 6}
\label{Tp6CTP}
\end{figure}

\begin{figure}
\centering
\includegraphics[width=0.4\textwidth]{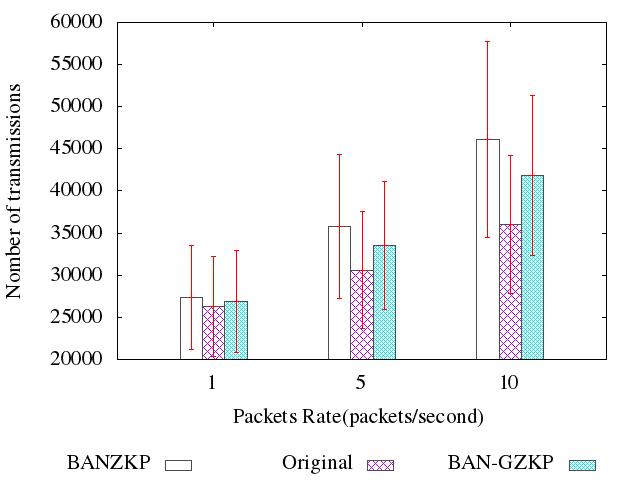}
\caption{Number of transmissions for CTP in posture 7}
\label{Tp7CTP}
\end{figure}

\begin{figure}
\centering
\includegraphics[width=0.4\textwidth]{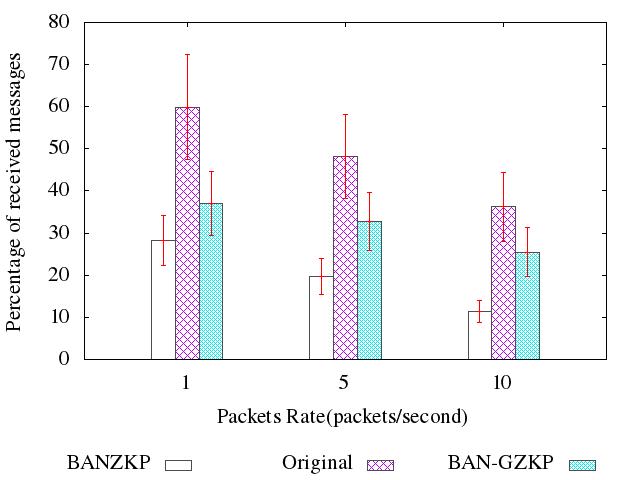}
\caption{Percentage of received messages for MiniAtt in posture  1}
\label{Rp1MiniAtt}
\end{figure}

\begin{figure}
\centering
\includegraphics[width=0.4\textwidth]{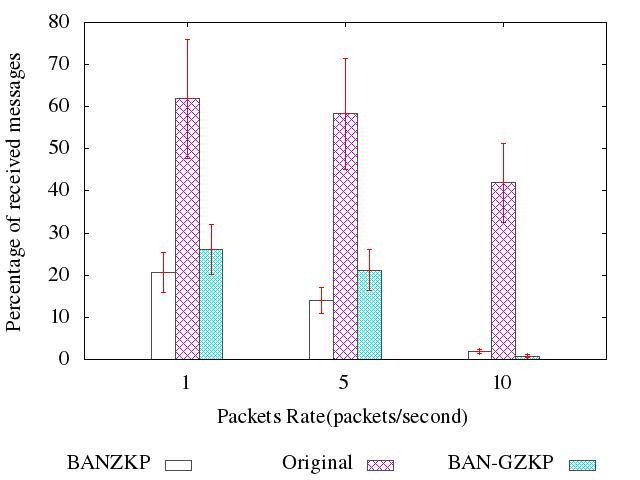}
\caption{Percentage of received messages for TreeBased in posture 6}
\label{Rp6TreeBased}
\end{figure}

We evaluated WBAN performance for each studied convergecast strategy under all seven mobility postures, in term of ratio of packet reception rate, data end-to-end delay and also numbers of transmissions plugged with BANZKP and BAN-GZKP, respectively. Figures from \ref{Rp1APAP} to \ref{Rp6TreeBased} show network performance in different strategies. In each figure, the red columns represent the original convergecast strategies without applying any authentication scheme; the white columns represent original strategies applying BANZKP; and the blue columns original strategies applying BAN-GZKP. The vertical bars of each column in figures refer to the confidence intervals. Results computed from 200 runs of simulations observation samples. The confidence intervals $\beta$ is calculated as:
\begin{align} 
	\beta = t(\alpha, R-1) \centerdot \frac{S}{\sqrt{R}}
\end{align}
where $\alpha$ is 0.005, means the confidence level is 95\%, $R$ is the number of observation samples and the $S$ is the standard deviation of observation samples.

For \emph{APAP} strategy (see Figures from \ref{Rp1APAP} to \ref{Tp7APAP}, a example for posture 1 Walking), by plugging ZKP scheme into the original APAP strategy, the WBAN performance decreases in general: lower ratio of packets reception, higher end-to-end delay and number of the transmissions, due to the additional ZKP authentication scheme added to the original one. When focusing on the comparison between BANZKP and BAN-GZKP, we noticed that BAN-GZKP has higher ratio of reception, lower end-to-end delay and number of transmissions in all the postures except the number of transmissions in the Posture 6 Sleeping, see Figure \ref{Tp6APAP}, where BANZKP has fewer number of transmissions than BAN-GZKP. In Posture 6, links between the nearby nodes to the sink may have important channel attenuations due to the obstruction and reflection of human body when sleeping \cite{bu2017total}. Links who are far away from the sink affected by lower attenuations comparing with links who are close to the sink. In the case of the hop-by-hop BAN-GZKP, data packet has a big chance to be lost when reaching these links close to the sink. That means BAN-GZKP may waste all transmissions of hop-by-hop authentication before the data is finally lost. However in the BANZKP case, there is no transmission wasted, since once the authentication packet is lost, after the timeout, the source node will notice the loss and close this transmission session. The BANZKP thus is better in terms of number of transmissions in Posture 6 Sleeping.


\begin{figure}
\centering
\includegraphics[width=0.4\textwidth]{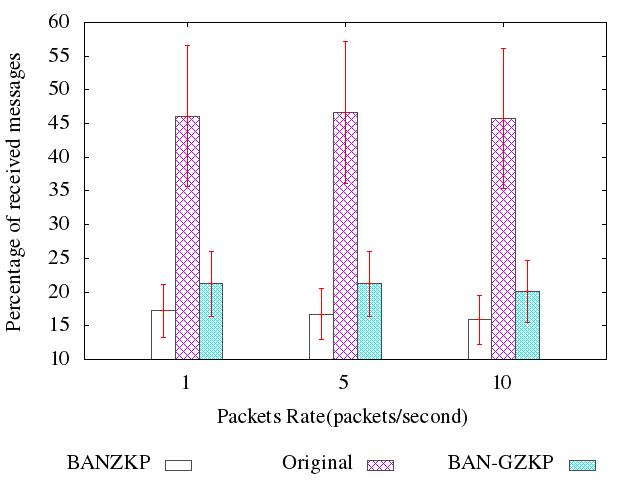}
\caption{Percentage of received messages for APAP in posture 2}
\label{run}
\end{figure}

\begin{figure}
\centering
\includegraphics[width=0.4\textwidth]{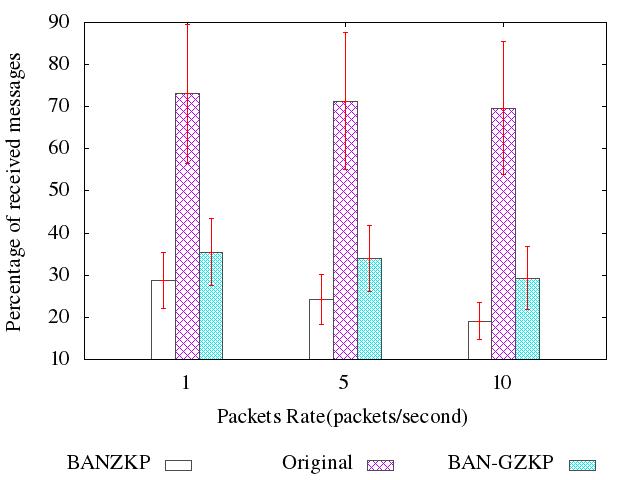}
\caption{Percentage of received messages for APAP in posture 3}
\label{weak}
\end{figure}

\begin{figure}
\centering
\includegraphics[width=0.4\textwidth]{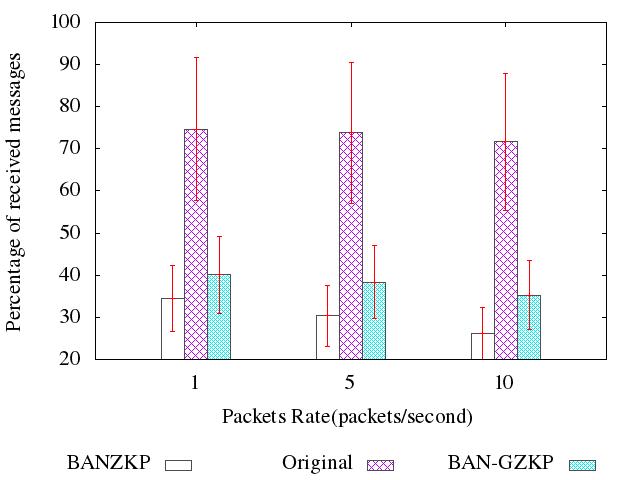}
\caption{Percentage of received messages for APAP in posture 4}
\label{sit}
\end{figure}

\begin{figure}
\centering
\includegraphics[width=0.4\textwidth]{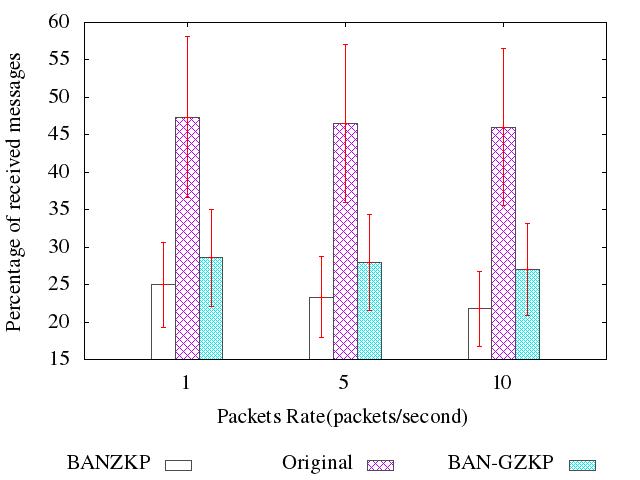}
\caption{Percentage of received messages for APAP in posture 5}
\label{lie}
\end{figure}

\begin{figure}
\centering
\includegraphics[width=0.4\textwidth]{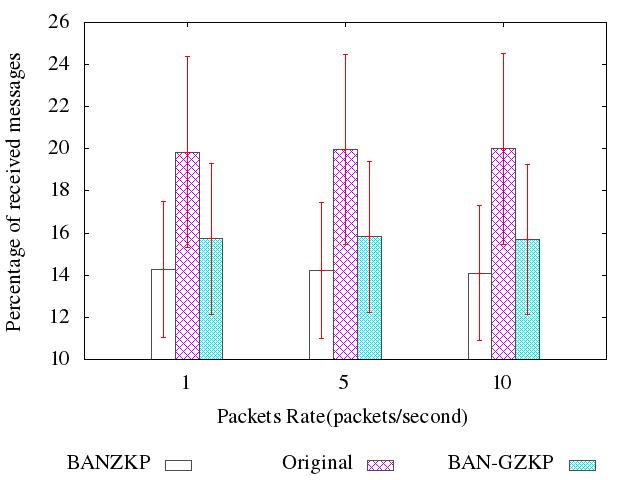}
\caption{Percentage of received messages for APAP in posture 6}
\label{sleep}
\end{figure}

\begin{figure}
\centering
\includegraphics[width=0.4\textwidth]{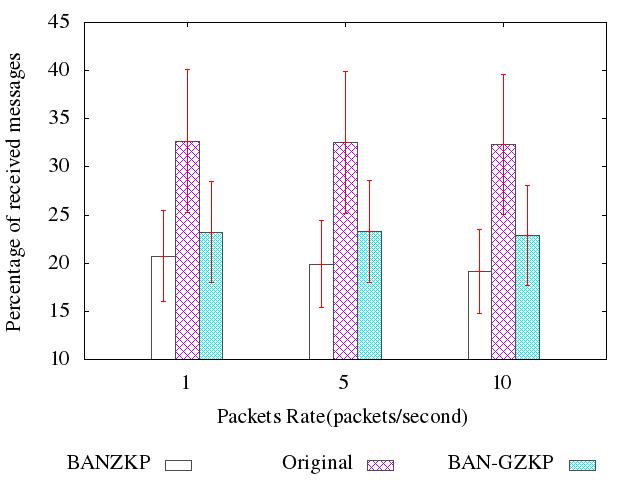}
\caption{Percentage of received messages for APAP in posture 7}
\label{jacket}
\end{figure}
\begin{figure}
\centering
\includegraphics[width=0.4\textwidth]{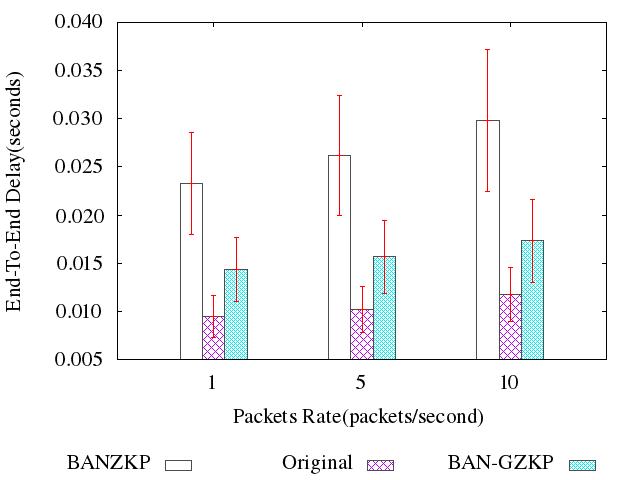}
\caption{End-To-End Delay for APAP in posture 1}
\label{walk}
\end{figure}

\begin{figure}
\centering
\includegraphics[width=0.4\textwidth]{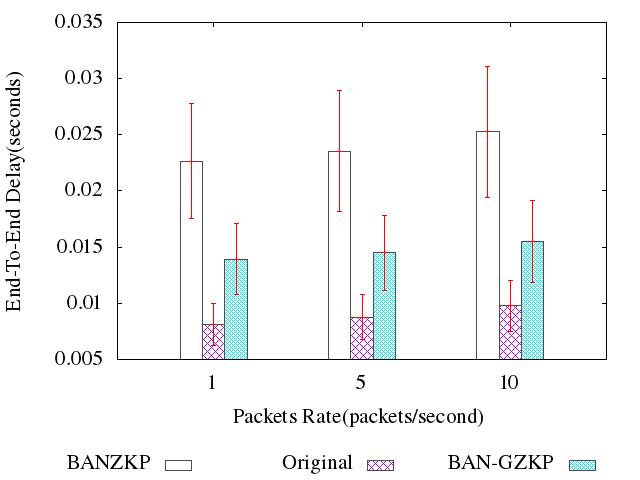}
\caption{End-To-End Delay for APAP in posture 2}
\label{run}
\end{figure}

\begin{figure}
\centering
\includegraphics[width=0.4\textwidth]{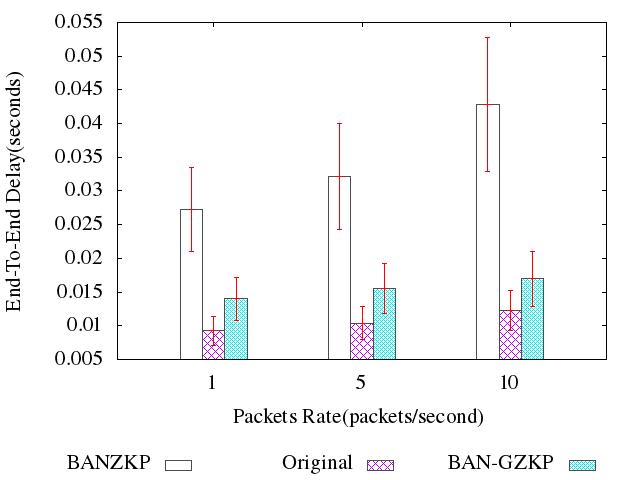}
\caption{End-To-End Delay for APAP in posture 3}
\label{weak}
\end{figure}

\begin{figure}
\centering
\includegraphics[width=0.4\textwidth]{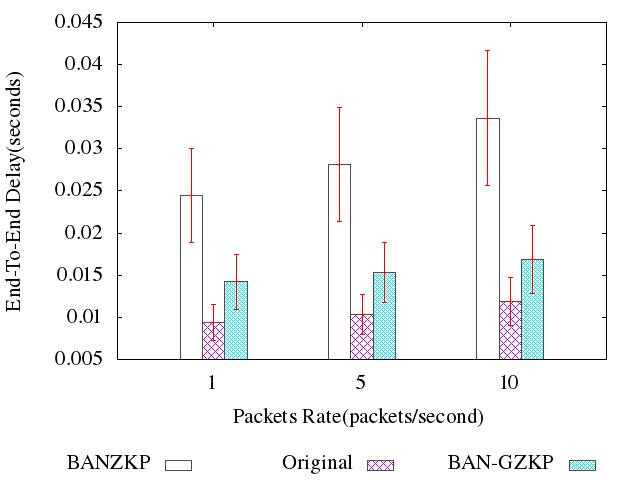}
\caption{End-To-End Delay for APAP in posture 4}
\label{sit}
\end{figure}

\begin{figure}
\centering
\includegraphics[width=0.4\textwidth]{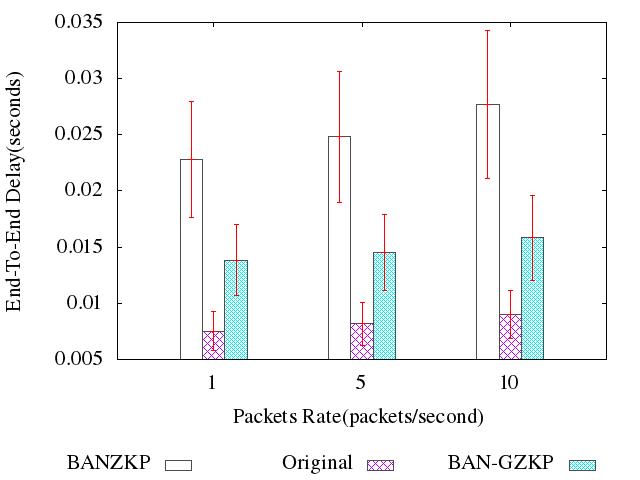}
\caption{End-To-End Delay for APAP in posture 5}
\label{lie}
\end{figure}

\begin{figure}
\centering
\includegraphics[width=0.4\textwidth]{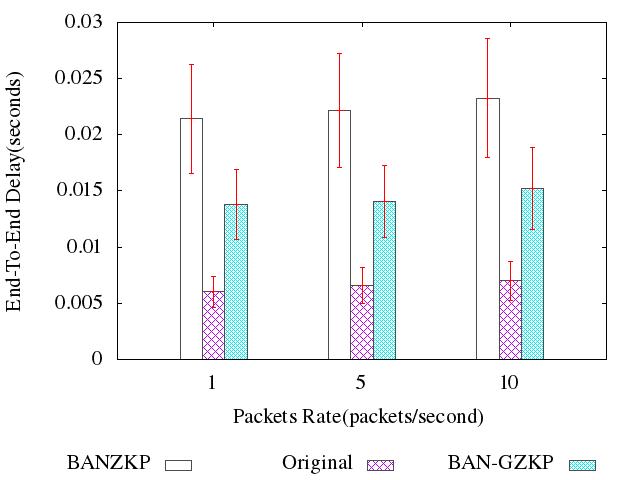}
\caption{End-To-End Delay for APAP in posture 6}
\label{sleep}
\end{figure}

\begin{figure}
\centering
\includegraphics[width=0.4\textwidth]{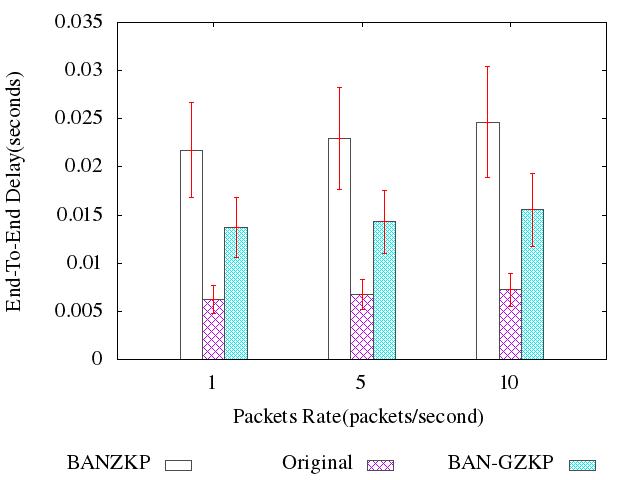}
\caption{End-To-End Delay for APAP in posture 7}
\label{jacket}
\end{figure}
\begin{figure}
\centering
\includegraphics[width=0.4\textwidth]{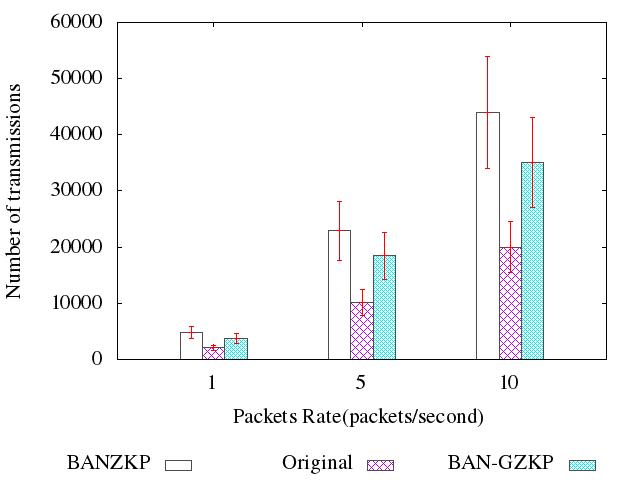}
\caption{Number of transmissions for APAP in posture 1}
\label{walk}
\end{figure}

\begin{figure}
\centering
\includegraphics[width=0.4\textwidth]{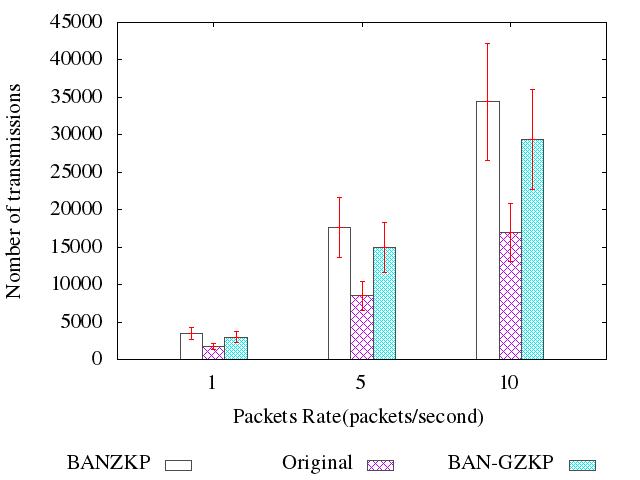}
\caption{Number of transmissions for APAP in posture 2}
\label{run}
\end{figure}

\begin{figure}
\centering
\includegraphics[width=0.4\textwidth]{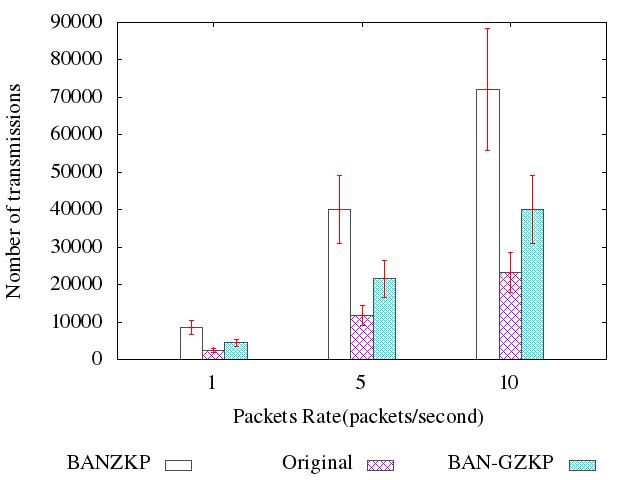}
\caption{Number of transmissions for APAP in posture 3}
\label{weak}
\end{figure}

\begin{figure}
\centering
\includegraphics[width=0.4\textwidth]{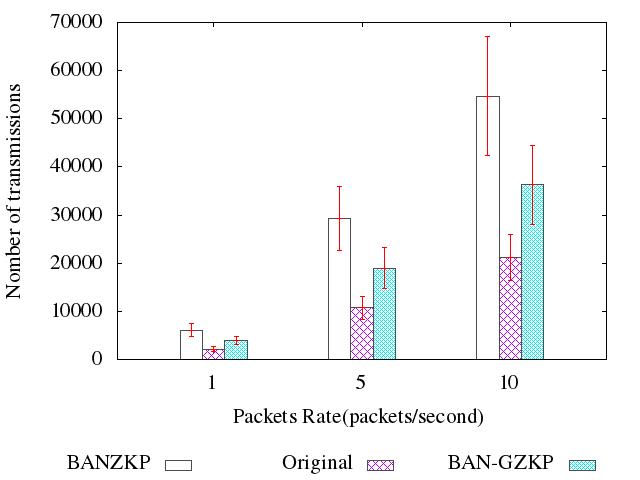}
\caption{Number of transmissions for APAP in posture 4}
\label{sit}
\end{figure}

\begin{figure}
\centering
\includegraphics[width=0.4\textwidth]{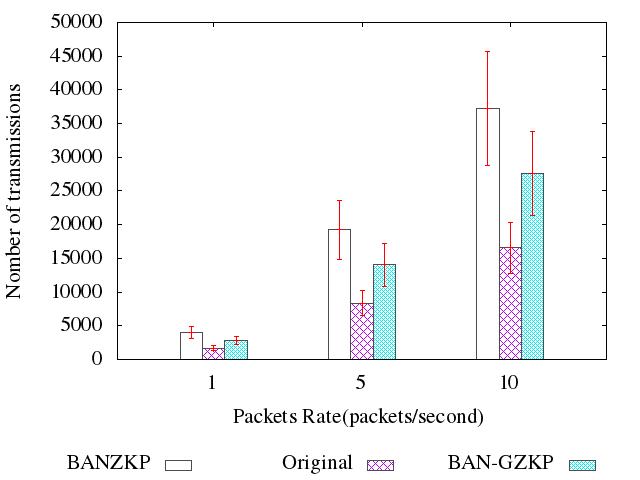}
\caption{Number of transmissions for APAP in posture 5}
\label{lie}
\end{figure}

For \emph{CTP} strategy (see Figures from \ref{Rp1CTP} to \ref{Tp7CTP}, a example for posture 1 Walking), we noticed that the performance parameters have the same behaviour as shown in strategy APAP. BAN-GZKP has better performances than BANZKP in terms of ratio of packets reception, end-to-end delay and number of transmissions except the number of transmissions in Posture 6 Sleeping, see Figure \ref{Tp6CTP}. The reason is the same as for the strategy APAP: BAN-GZKP may waste more transmissions than BANZKP before a packet gets lost if the link error occurs near the sink. The reason why BANZKP and BAN-GZKP have the same behaviour in both CTP and APAP, respectively is as follows: CTP constructs a tree from each source node to the sink at the initial phase and keeps updating the tree during the transmission; APAP on the other side uses and keeps pre-setted multiple-reception paths. The common point is that both CTP and APAP tend to maintain good routes from sources to the sink. When a node has a packet to send, it can directly send the packets to an already-keep-in-mid next hop.

\begin{figure}
\centering
\includegraphics[width=0.4\textwidth]{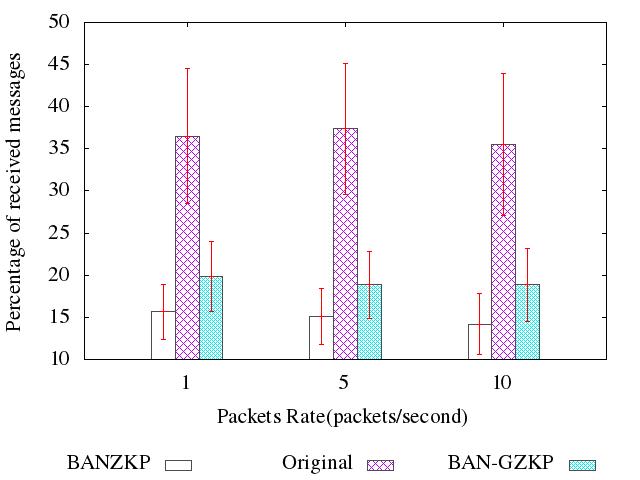}
\caption{Percentage of received messages for CTP in posture 2}
\label{run}
\end{figure}

\begin{figure}
\centering
\includegraphics[width=0.4\textwidth]{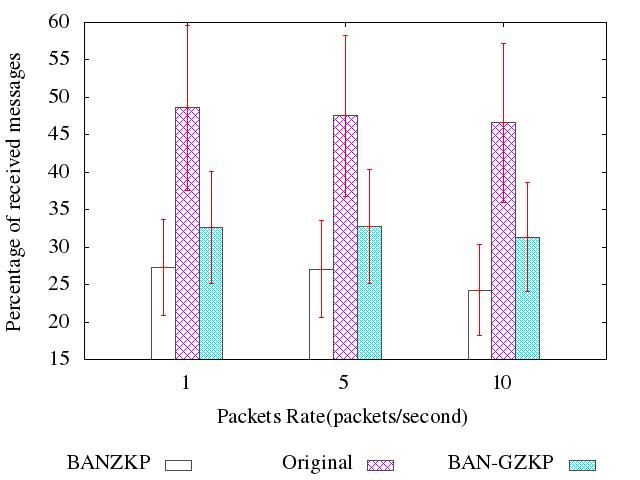}
\caption{Percentage of received messages for CTP in posture 3}
\label{weak}
\end{figure}

\begin{figure}
\centering
\includegraphics[width=0.4\textwidth]{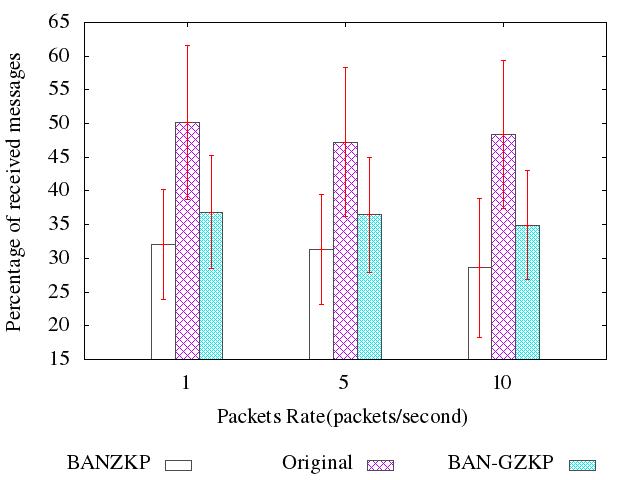}
\caption{Percentage of received messages for CTP in posture 4}
\label{sit}
\end{figure}

\begin{figure}
\centering
\includegraphics[width=0.4\textwidth]{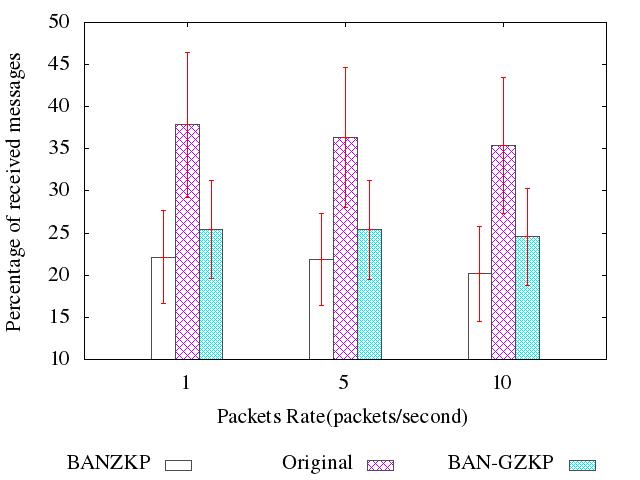}
\caption{Percentage of received messages for CTP in posture 5}
\label{lie}
\end{figure}

\begin{figure}
\centering
\includegraphics[width=0.4\textwidth]{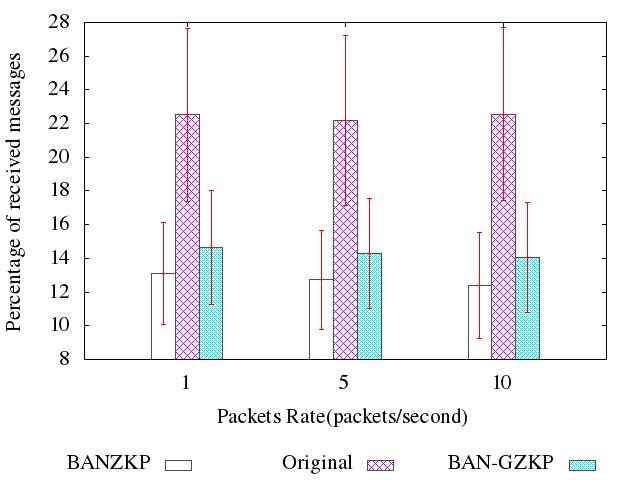}
\caption{Percentage of received messages for CTP in posture 6}
\label{sleep}
\end{figure}

\begin{figure}
\centering
\includegraphics[width=0.4\textwidth]{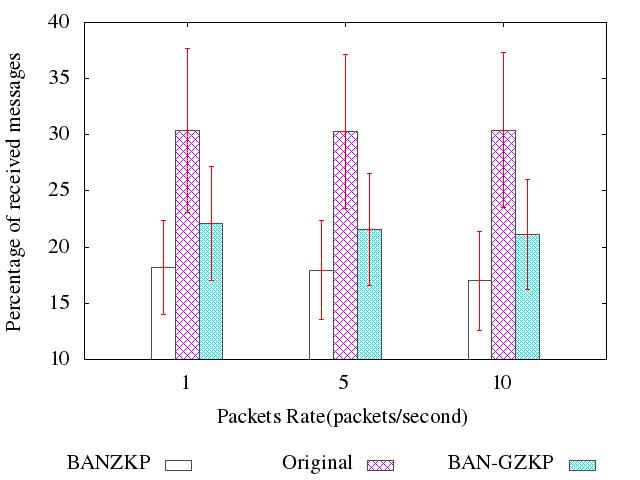}
\caption{Percentage of received messages for CTP in posture 7}
\label{jacket}
\end{figure}
\begin{figure}
\centering
\includegraphics[width=0.4\textwidth]{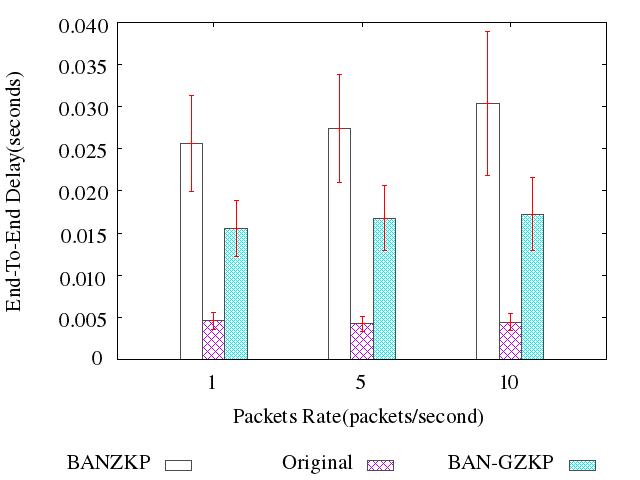}
\caption{End-To-End Delay for CTP in posture 1}
\label{walk}
\end{figure}

\begin{figure}
\centering
\includegraphics[width=0.4\textwidth]{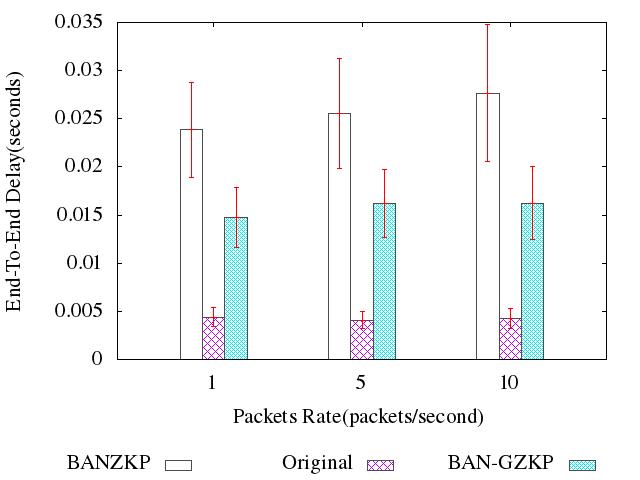}
\caption{End-To-End Delay for CTP in posture 2}
\label{run}
\end{figure}

\begin{figure}
\centering
\includegraphics[width=0.4\textwidth]{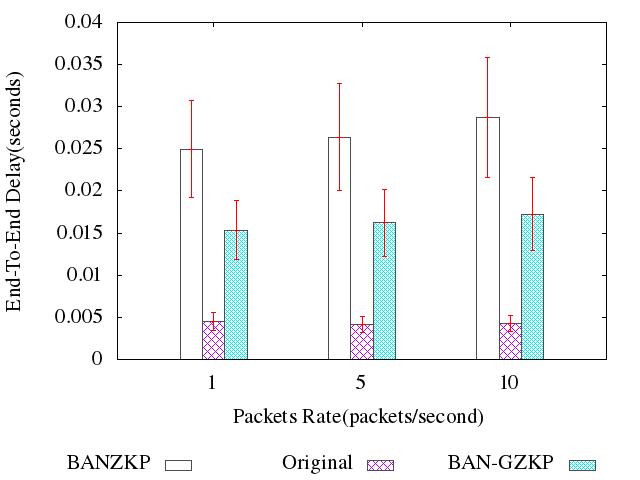}
\caption{End-To-End Delay for CTP in posture 3}
\label{weak}
\end{figure}

\begin{figure}
\centering
\includegraphics[width=0.4\textwidth]{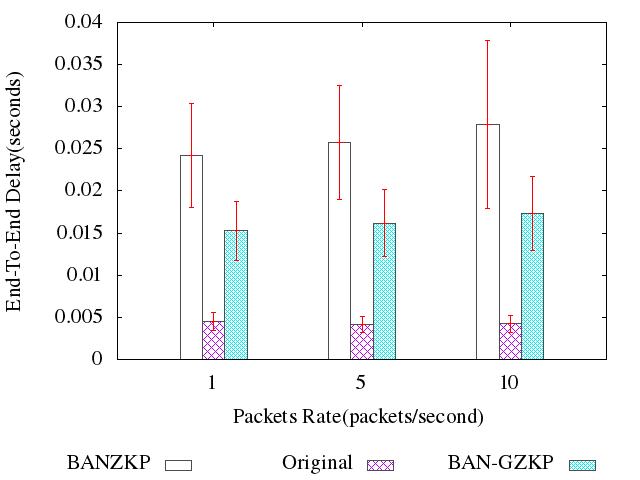}
\caption{End-To-End Delay for CTP in posture 4}
\label{sit}
\end{figure}

\begin{figure}
\centering
\includegraphics[width=0.4\textwidth]{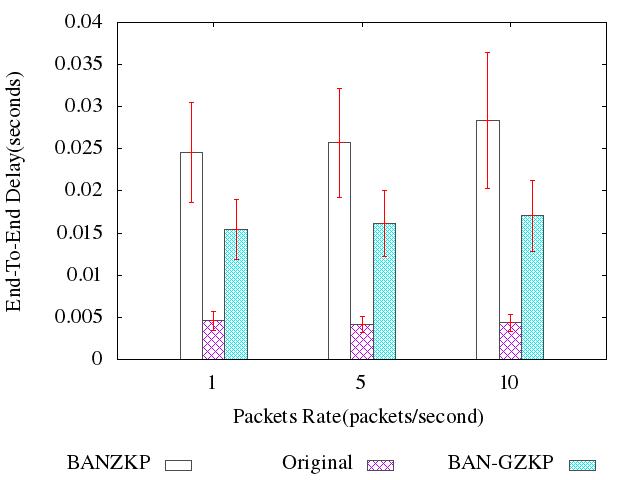}
\caption{End-To-End Delay for CTP in posture 5}
\label{lie}
\end{figure}

\begin{figure}
\centering
\includegraphics[width=0.4\textwidth]{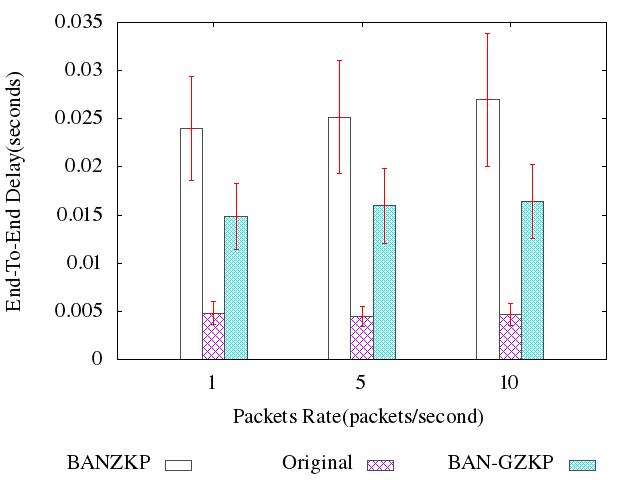}
\caption{End-To-End Delay for CTP in posture 6}
\label{sleep}
\end{figure}

\begin{figure}
\centering
\includegraphics[width=0.4\textwidth]{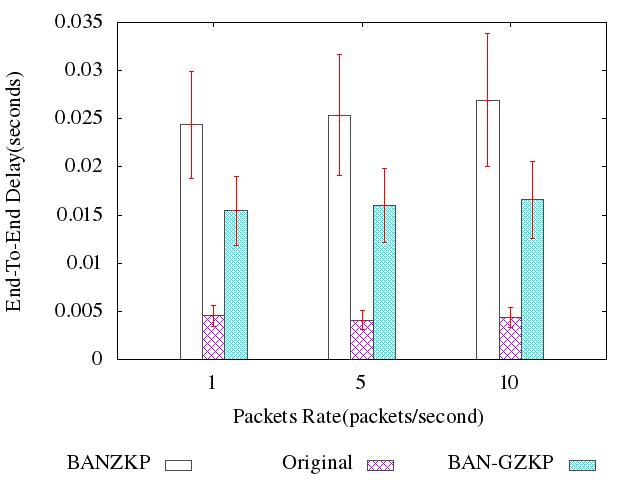}
\caption{End-To-End Delay for CTP in posture 7}
\label{jacket}
\end{figure}
\begin{figure}
\centering
\includegraphics[width=0.4\textwidth]{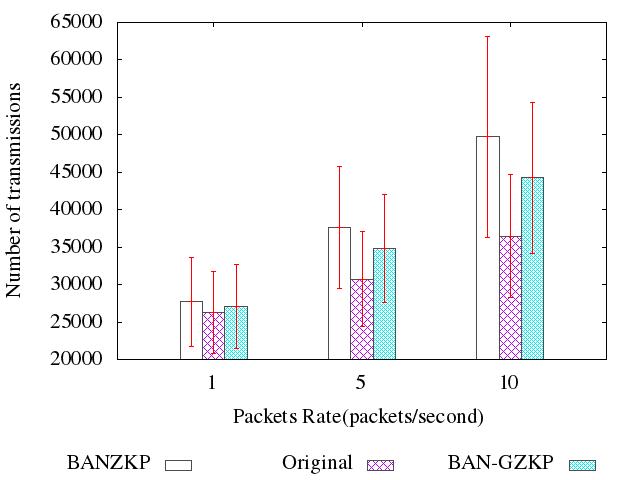}
\caption{Number of transmissions for CTP in posture 1}
\label{walk}
\end{figure}

\begin{figure}
\centering
\includegraphics[width=0.4\textwidth]{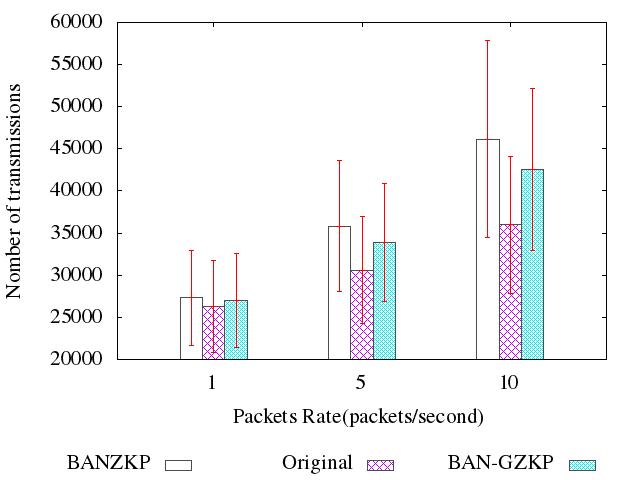}
\caption{Number of transmissions for CTP in posture 2}
\label{run}
\end{figure}

\begin{figure}
\centering
\includegraphics[width=0.4\textwidth]{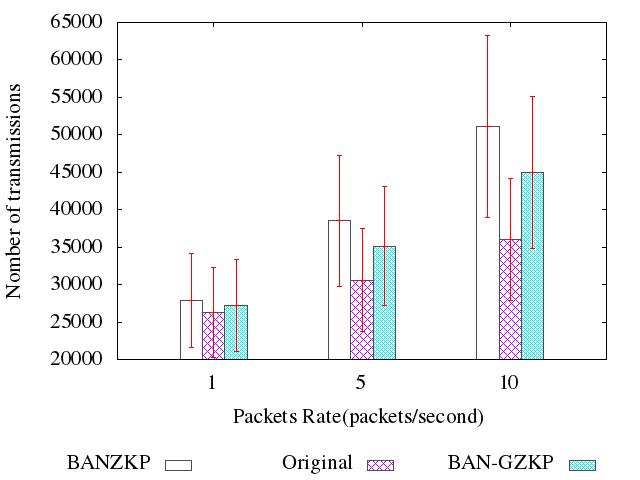}
\caption{Number of transmissions for CTP in posture 3}
\label{weak}
\end{figure}

\begin{figure}
\centering
\includegraphics[width=0.4\textwidth]{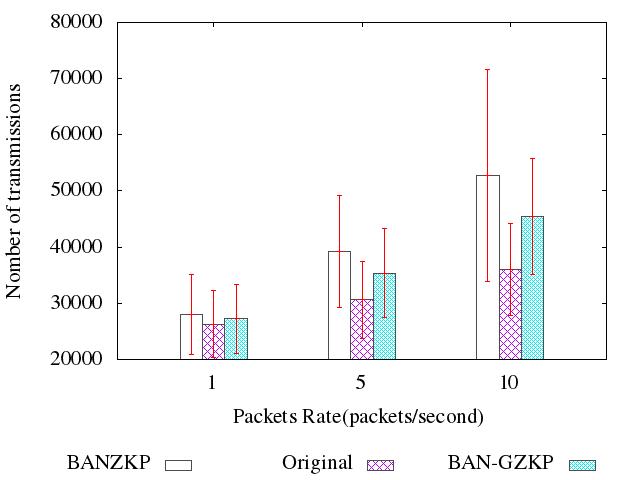}
\caption{Number of transmissions for CTP in posture 4}
\label{sit}
\end{figure}

\begin{figure}
\centering
\includegraphics[width=0.4\textwidth]{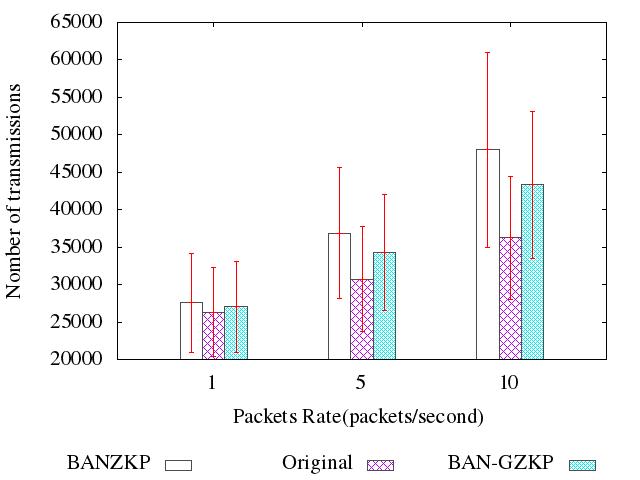}
\caption{Number of transmissions for CTP in posture 5}
\label{lie}
\end{figure}

\begin{figure}
\centering
\includegraphics[width=0.4\textwidth]{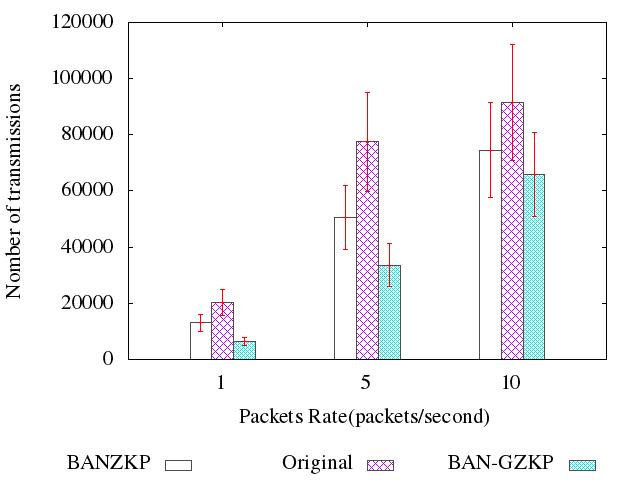}
\caption{Number of transmissions for MiniAtt in posture 7}
\label{Tp7MiniAtt}
\end{figure}

For \emph{MiniAtt} strategy (see Figures from \ref{Rp1MiniAtt} to \ref{Tp7MiniAtt}, a example for posture 1 Walking), we get lower end-to-end delay and number of transmissions when using BANZKP and BAN-GZKP than the original MiniAtt strategy. That is due to the negotiation nature of MiniAtt strategy: before any transmission, sender has to ask around its neighbours, who could have the smallest attenuation to the sink. That makes data packets generated by nodes far away from sink need more time and more transmissions to reach the sink than data packets generated by nodes close to the sink. The reception of distant packets extends the average of end-to-end delay and the average of number of transmissions comparing with the original strategy. When plugged with the BANZKP and BAN-GZKP, the additional ZKP message exchange makes it harder for distant packets to reach the sink. As the distant packets occurs fewer percentage of the total packets reception, the average of end-to-end delay and number of transmissions can thus reduce. When comparing BANZKP and BAN-GZKP, in the case of MiniAtt, BAN-GZKP always has better ratio of packets reception and number of transmissions. Additionally it has comparable end-to-end delay in all the postures.

\begin{figure}
\centering
\includegraphics[width=0.4\textwidth]{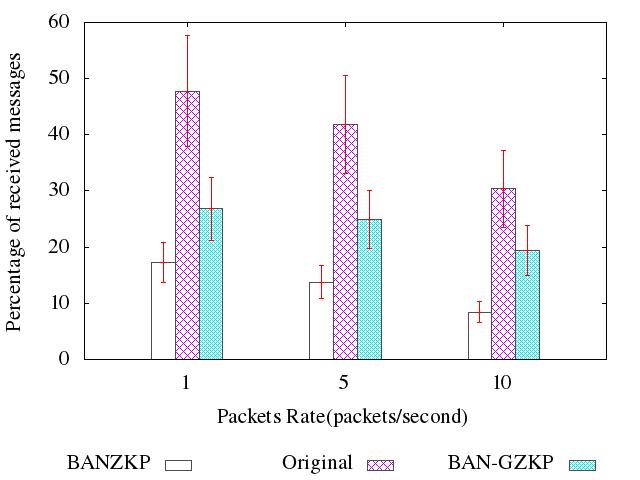}
\caption{Percentage of received messages for MiniAtt in posture 2}
\label{run}
\end{figure}

\begin{figure}
\centering
\includegraphics[width=0.4\textwidth]{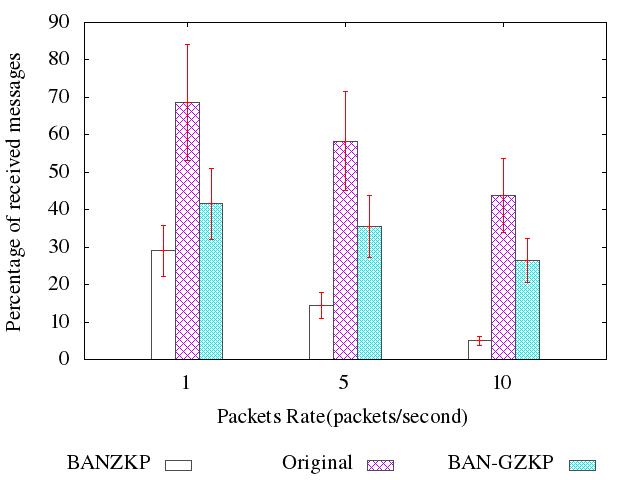}
\caption{Percentage of received messages for MiniAtt in posture 3}
\label{weak}
\end{figure}

\begin{figure}
\centering
\includegraphics[width=0.4\textwidth]{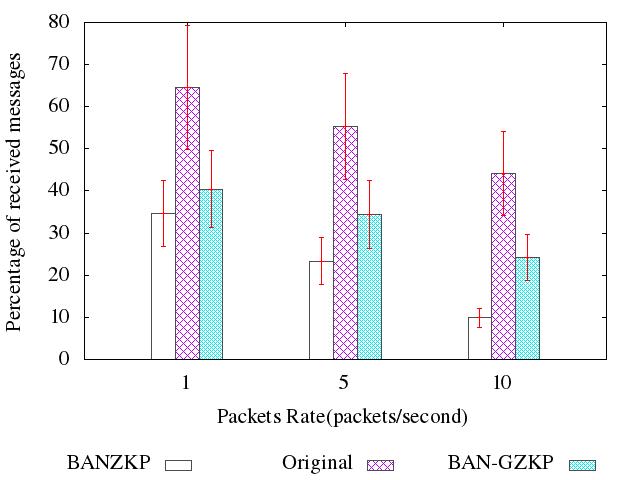}
\caption{Percentage of received messages for MiniAtt in posture 4}
\label{sit}
\end{figure}

\begin{figure}
\centering
\includegraphics[width=0.4\textwidth]{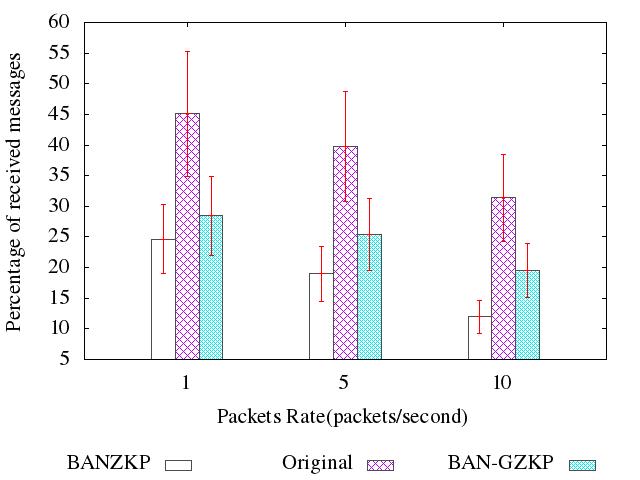}
\caption{Percentage of received messages for MiniAtt in posture 5}
\label{lie}
\end{figure}

\begin{figure}
\centering
\includegraphics[width=0.4\textwidth]{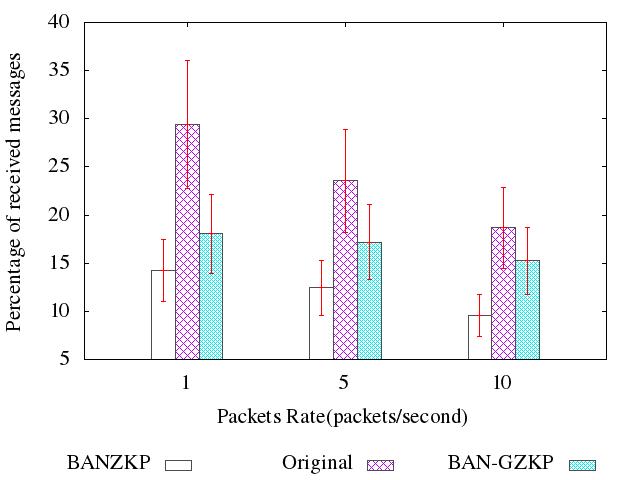}
\caption{Percentage of received messages for MiniAtt in posture 6}
\label{sleep}
\end{figure}

\begin{figure}
\centering
\includegraphics[width=0.4\textwidth]{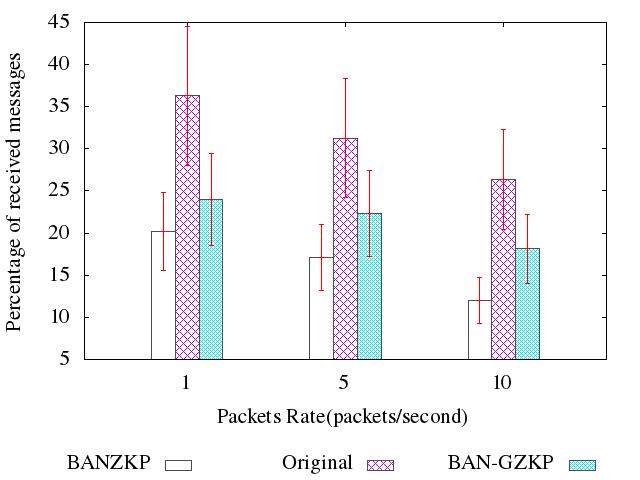}
\caption{Percentage of received messages for MiniAtt in posture 7}
\label{jacket}
\end{figure}
\begin{figure}
\centering
\includegraphics[width=0.4\textwidth]{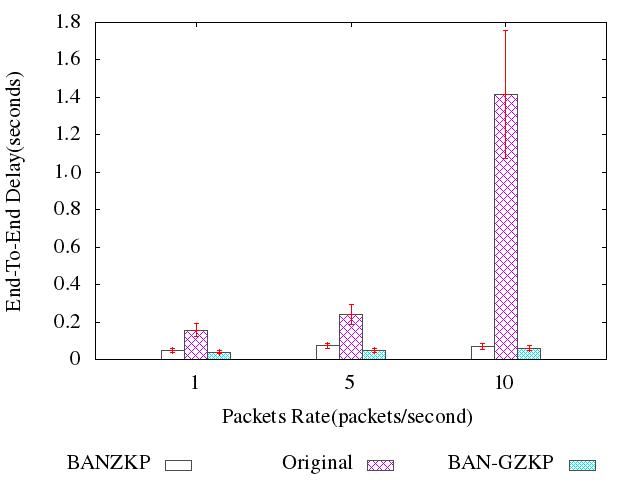}
\caption{End-To-End Delay for MiniAtt in posture 1}
\label{walk}
\end{figure}

\begin{figure}
\centering
\includegraphics[width=0.4\textwidth]{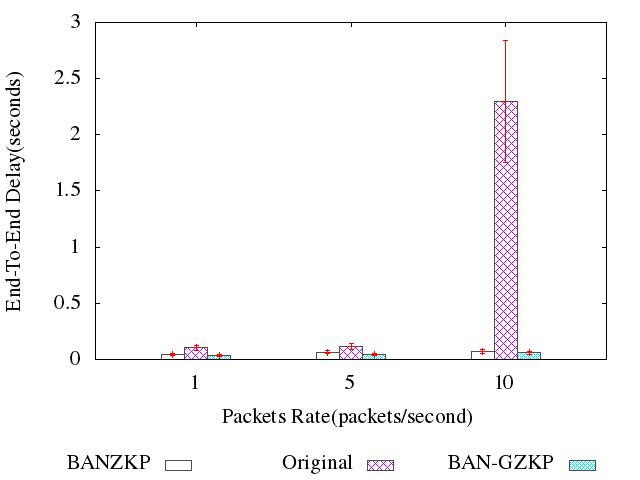}
\caption{End-To-End Delay for MiniAtt in posture 2}
\label{run}
\end{figure}

\begin{figure}
\centering
\includegraphics[width=0.4\textwidth]{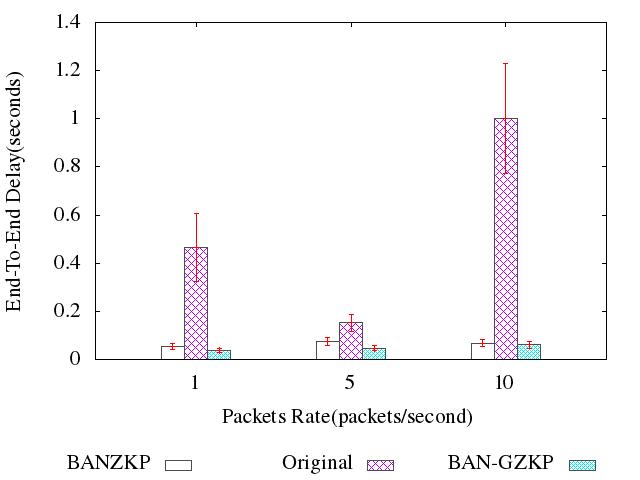}
\caption{End-To-End Delay for MiniAtt in posture 3}
\label{weak}
\end{figure}

\begin{figure}
\centering
\includegraphics[width=0.4\textwidth]{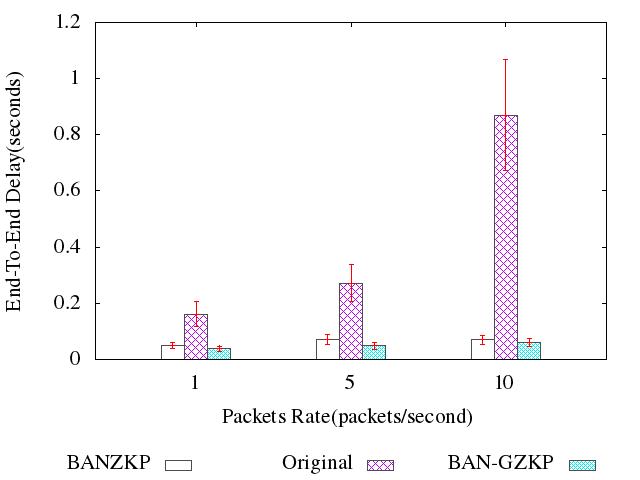}
\caption{End-To-End Delay for MiniAtt in posture 4}
\label{sit}
\end{figure}

\begin{figure}
\centering
\includegraphics[width=0.4\textwidth]{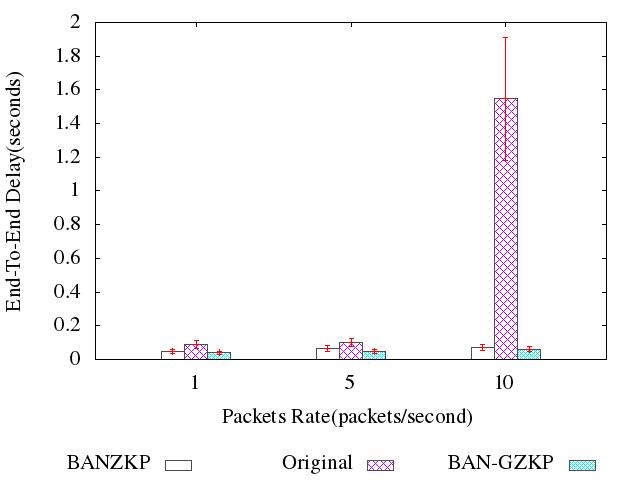}
\caption{End-To-End Delay for MiniAtt in posture 5}
\label{lie}
\end{figure}

\begin{figure}
\centering
\includegraphics[width=0.4\textwidth]{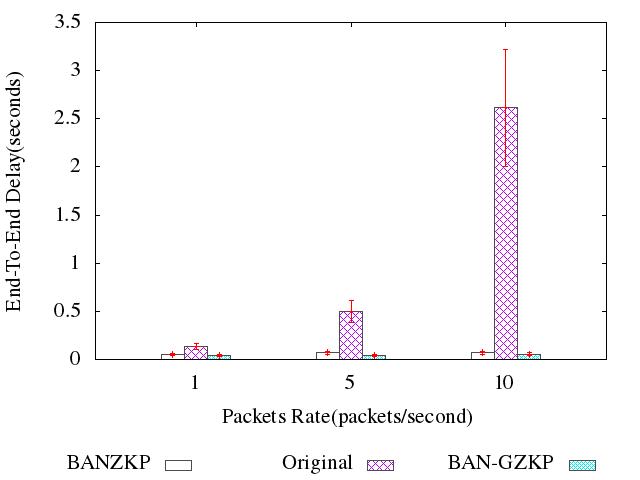}
\caption{End-To-End Delay for MiniAtt in posture 6}
\label{sleep}
\end{figure}

\begin{figure}
\centering
\includegraphics[width=0.4\textwidth]{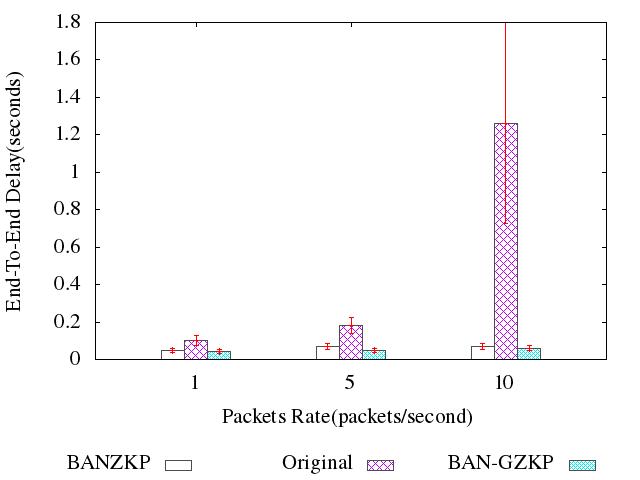}
\caption{End-To-End Delay for MiniAtt in posture 7}
\label{jacket}
\end{figure}
\begin{figure}
\centering
\includegraphics[width=0.4\textwidth]{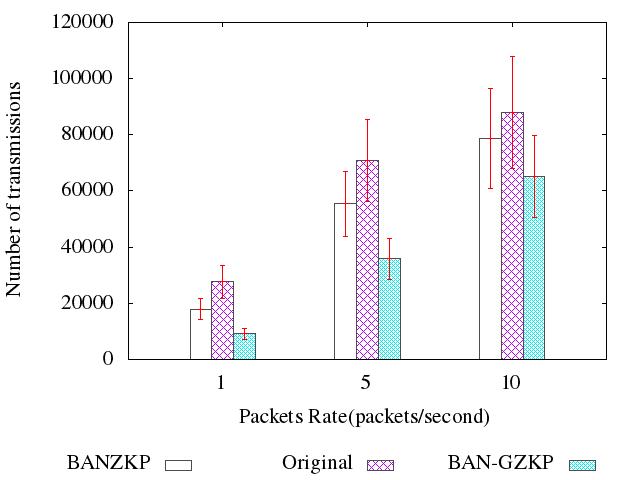}
\caption{Number of transmissions for MiniAtt in posture 1}
\label{walk}
\end{figure}

\begin{figure}
\centering
\includegraphics[width=0.4\textwidth]{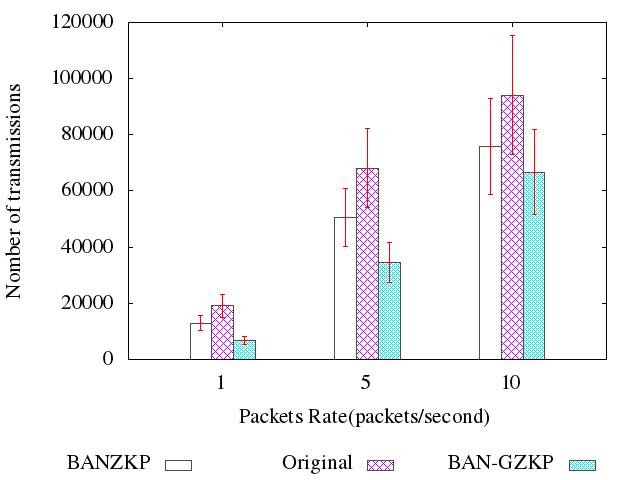}
\caption{Number of transmissions for MiniAtt in posture 2}
\label{run}
\end{figure}

\begin{figure}
\centering
\includegraphics[width=0.4\textwidth]{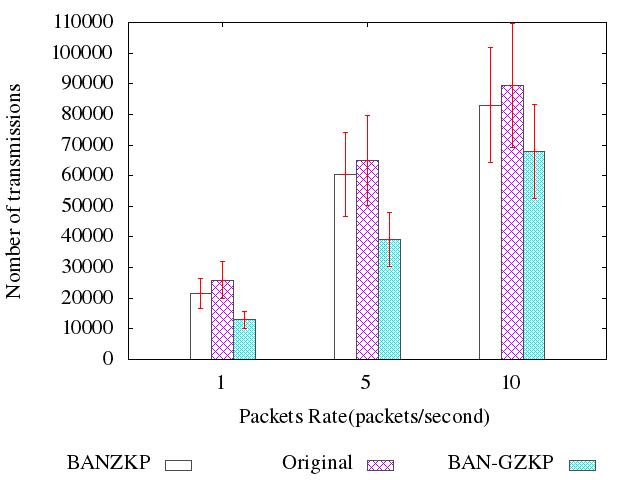}
\caption{Number of transmissions for MiniAtt in posture 3}
\label{weak}
\end{figure}

\begin{figure}
\centering
\includegraphics[width=0.4\textwidth]{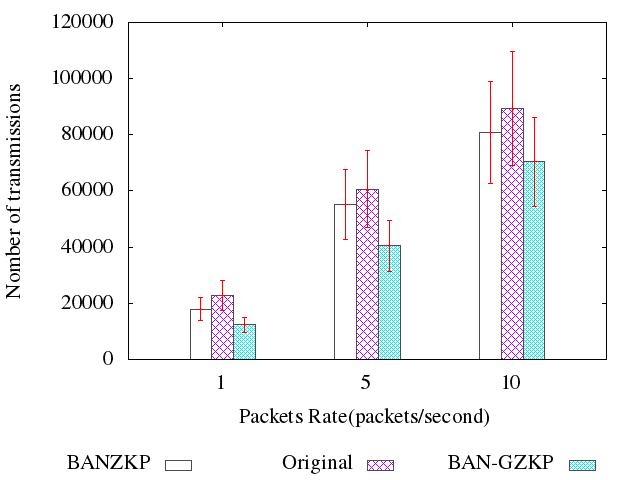}
\caption{Number of transmissions for MiniAtt in posture 4}
\label{sit}
\end{figure}

\begin{figure}
\centering
\includegraphics[width=0.4\textwidth]{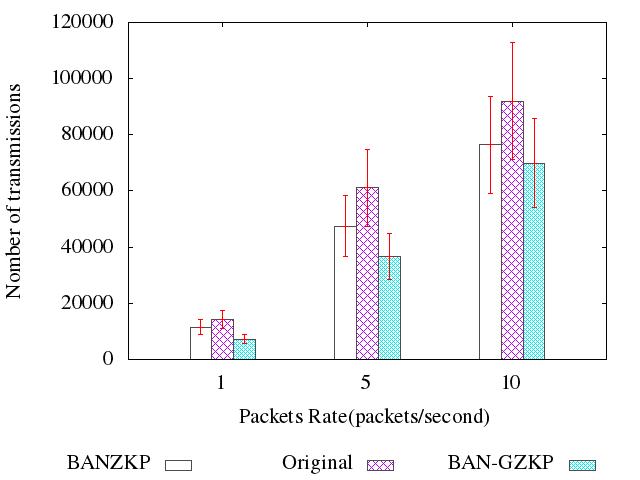}
\caption{Number of transmissions for MiniAtt in posture 5}
\label{lie}
\end{figure}

\begin{figure}
\centering
\includegraphics[width=0.4\textwidth]{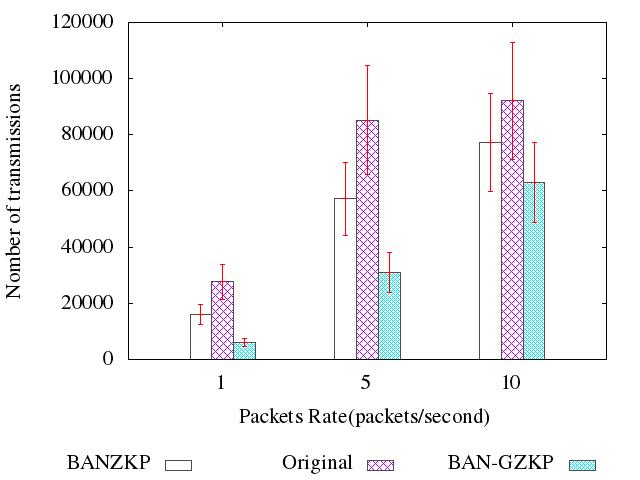}
\caption{Number of transmissions for MiniAtt in posture 6}
\label{sleep}
\end{figure}

\begin{figure}
\centering
\includegraphics[width=0.4\textwidth]{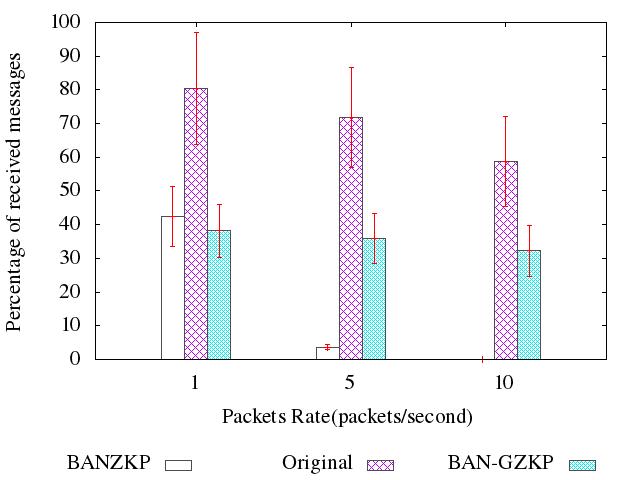}
\caption{Percentage of received messages for FloodToSink in posture  1}
\label{Rp1FloodToSink}
\end{figure}

\begin{figure}
\centering
\includegraphics[width=0.4\textwidth]{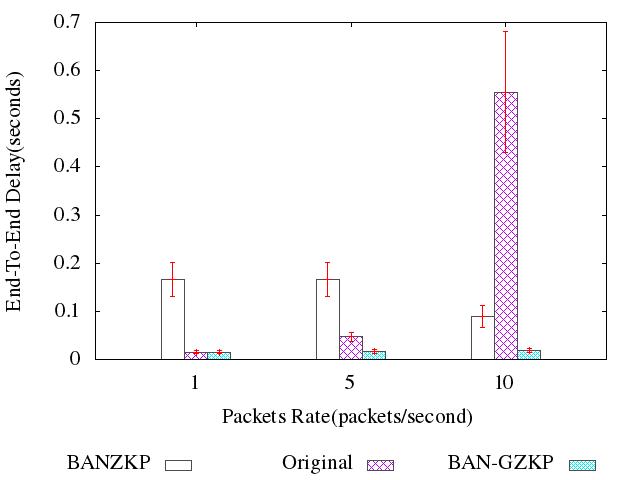}
\caption{End-To-End Delay for FloodToSink in posture 1}
\label{Dp1FloodToSink}
\end{figure}

\begin{figure}
\centering
\includegraphics[width=0.4\textwidth]{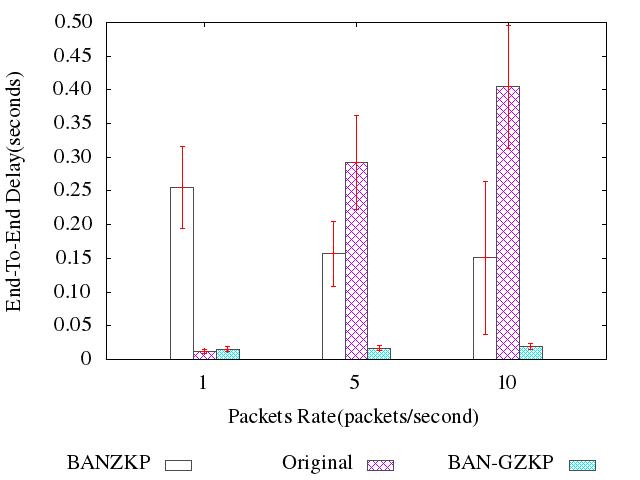}
\caption{End-To-End Delay for FloodToSink in posture 3}
\label{Dp3FloodToSink}
\end{figure}

\begin{figure}
\centering
\includegraphics[width=0.4\textwidth]{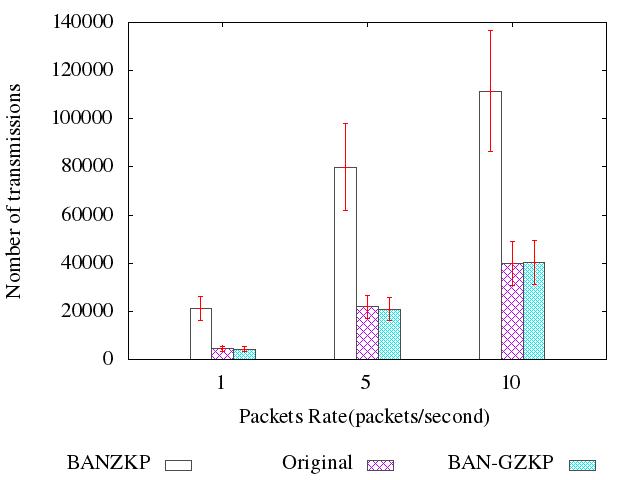}
\caption{Number of transmissions for FloodToSink in posture 7}
\label{Tp7FloodToSink}
\end{figure}

For \emph{FloodToSink} strategy (see Figures from \ref{Rp1FloodToSink} to \ref{Tp7FloodToSink}, a example for posture 1 Walking), in general, BAN-GZKP is better than BANZKP, even better than original FloodToSink strategy in terms of number of transmissions. That is because the hop-by-hop BAN-GZKP can limit the flooding transmissions all over the network. In terms of end-to-end delay, BANZKP and the original FloodToSink has varying behaviours in different postures, see Figures \ref{Dp1FloodToSink} and \ref{Dp3FloodToSink}: big variance of end-to-end delay, due to its broadcast nature. In terms of ratio of packets reception, original FloodToSink is better when using any ZKP scheme. However the ratio of the packets receptions decreases much faster in the case of BANZKP and original FloodToSink than in BAN-GZKP who is relatively stable. That is because with the increase of the packets generation rate, the number of the packets sent to the network will increase exponentially and lead to the network congestion.

\begin{figure}
\centering
\includegraphics[width=0.4\textwidth]{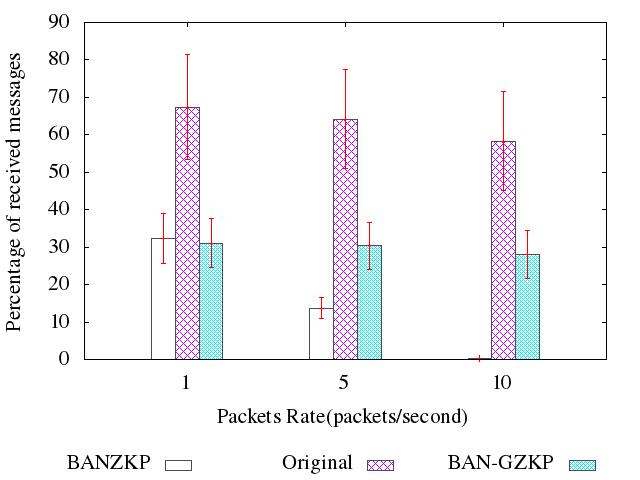}
\caption{Percentage of received messages for FloodToSink in posture 2}
\label{run}
\end{figure}

\begin{figure}
\centering
\includegraphics[width=0.4\textwidth]{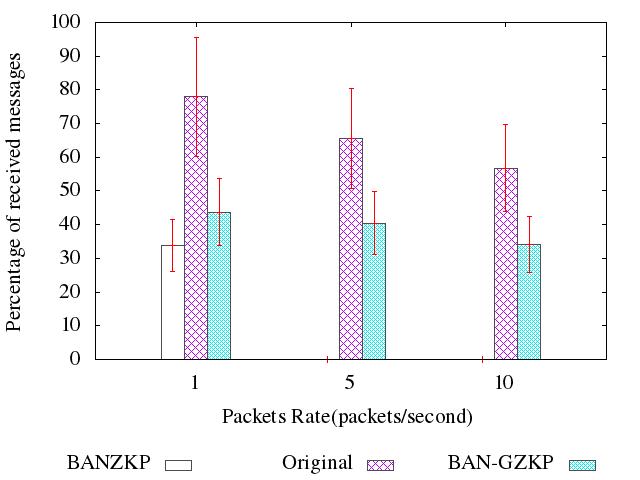}
\caption{Percentage of received messages for FloodToSink in posture 3}
\label{weak}
\end{figure}

\begin{figure}
\centering
\includegraphics[width=0.4\textwidth]{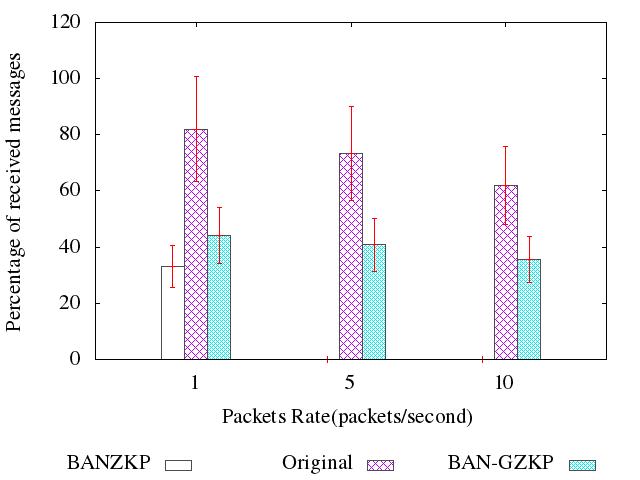}
\caption{Percentage of received messages for FloodToSink in posture 4}
\label{sit}
\end{figure}

\begin{figure}
\centering
\includegraphics[width=0.4\textwidth]{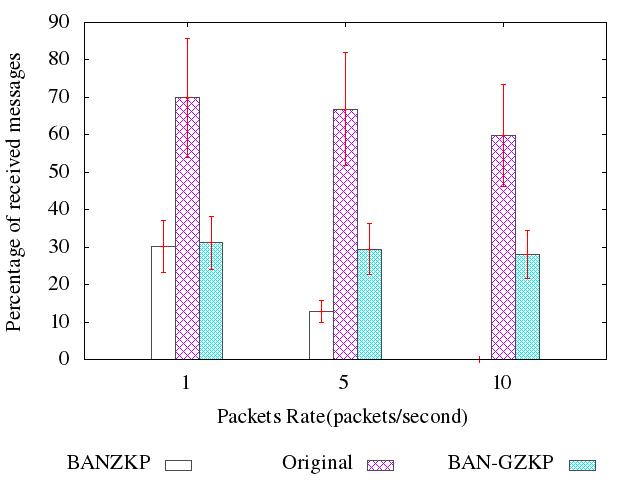}
\caption{Percentage of received messages for FloodToSink in posture 5}
\label{lie}
\end{figure}

\begin{figure}
\centering
\includegraphics[width=0.4\textwidth]{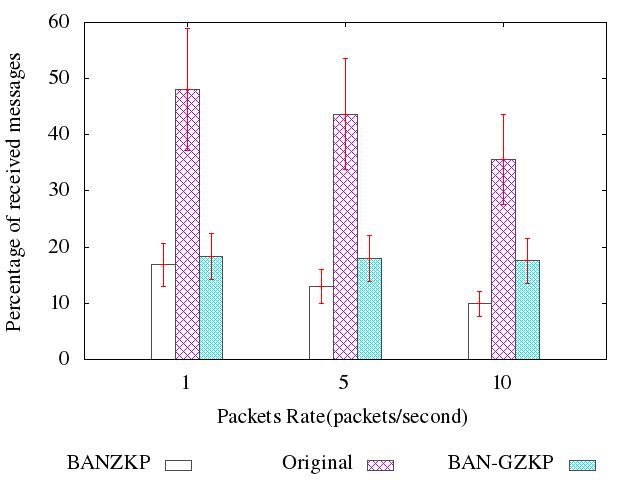}
\caption{Percentage of received messages for FloodToSink in posture 6}
\label{sleep}
\end{figure}

\begin{figure}
\centering
\includegraphics[width=0.4\textwidth]{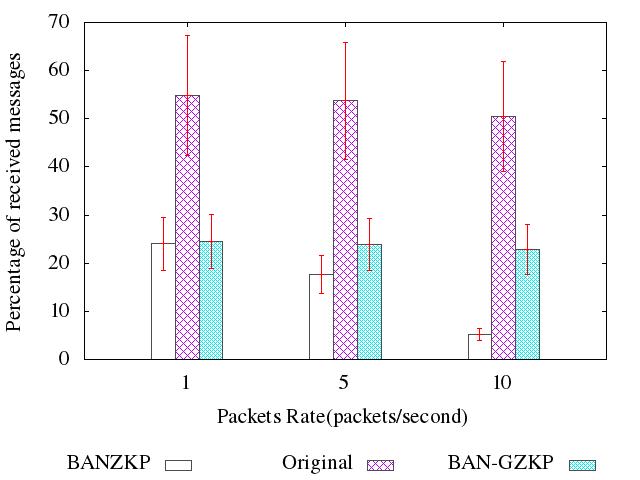}
\caption{Percentage of received messages for FloodToSink in posture 7}
\label{jacket}
\end{figure}

\begin{figure}
\centering
\includegraphics[width=0.4\textwidth]{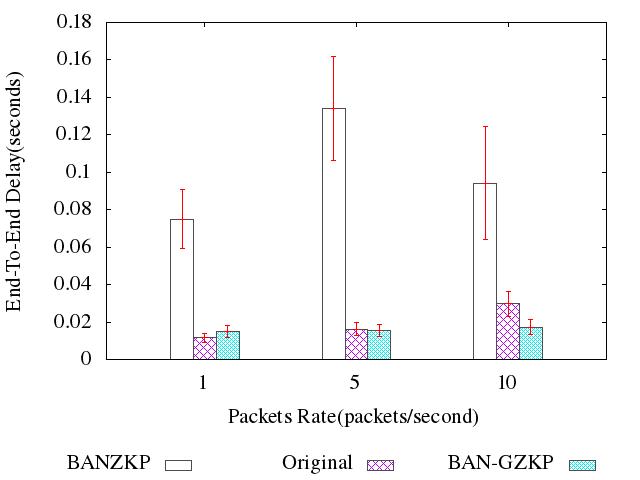}
\caption{End-To-End Delay for FloodToSink in posture 2}
\label{run}
\end{figure}

\begin{figure}
\centering
\includegraphics[width=0.4\textwidth]{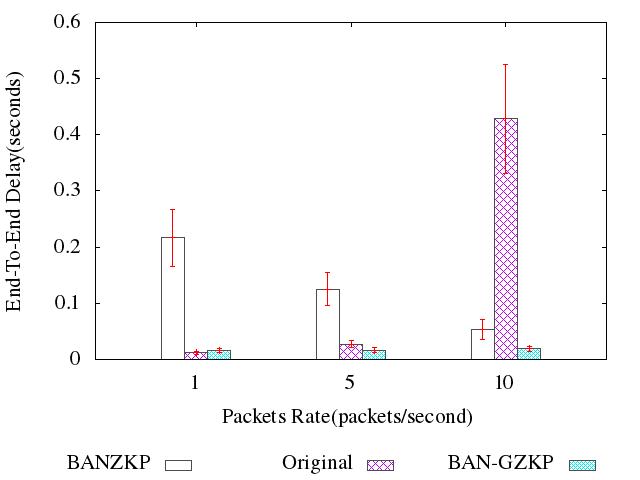}
\caption{End-To-End Delay for FloodToSink in posture 4}
\label{sit}
\end{figure}

\begin{figure}
\centering
\includegraphics[width=0.4\textwidth]{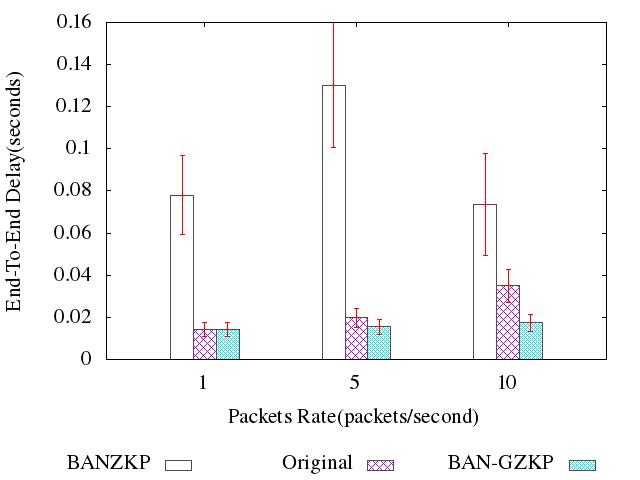}
\caption{End-To-End Delay for FloodToSink in posture 5}
\label{lie}
\end{figure}

\begin{figure}
\centering
\includegraphics[width=0.4\textwidth]{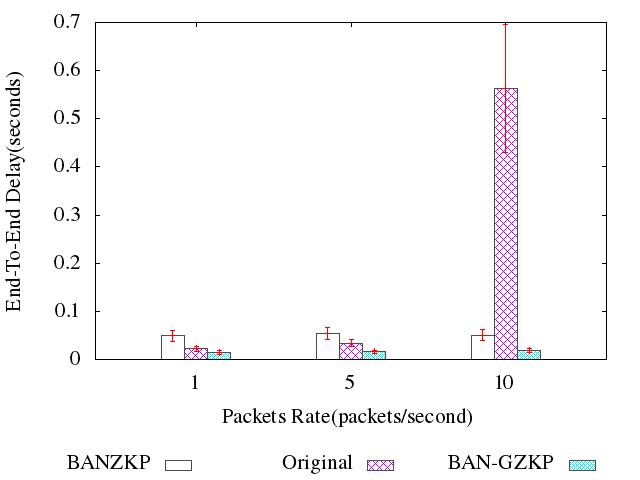}
\caption{End-To-End Delay for FloodToSink in posture 6}
\label{sleep}
\end{figure}

\begin{figure}
\centering
\includegraphics[width=0.4\textwidth]{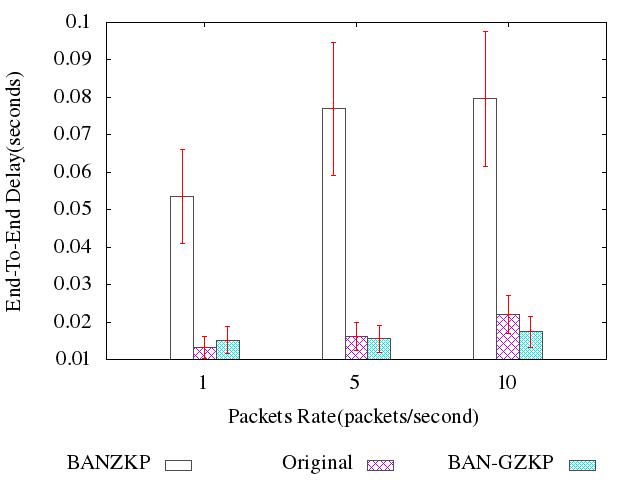}
\caption{End-To-End Delay for FloodToSink in posture 7}
\label{jacket}
\end{figure}
\begin{figure}
\centering
\includegraphics[width=0.4\textwidth]{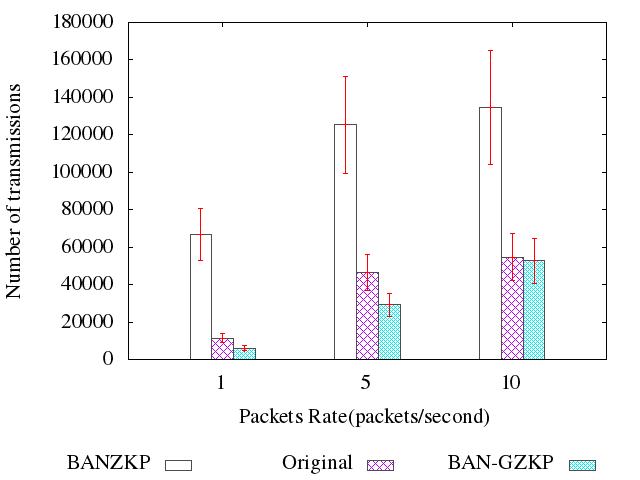}
\caption{Number of transmissions for FloodToSink in posture 1}
\label{walk}
\end{figure}

\begin{figure}
\centering
\includegraphics[width=0.4\textwidth]{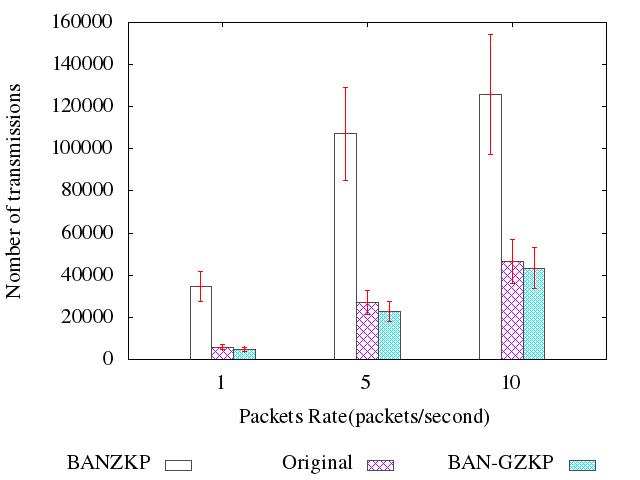}
\caption{Number of transmissions for FloodToSink in posture 2}
\label{run}
\end{figure}

\begin{figure}
\centering
\includegraphics[width=0.4\textwidth]{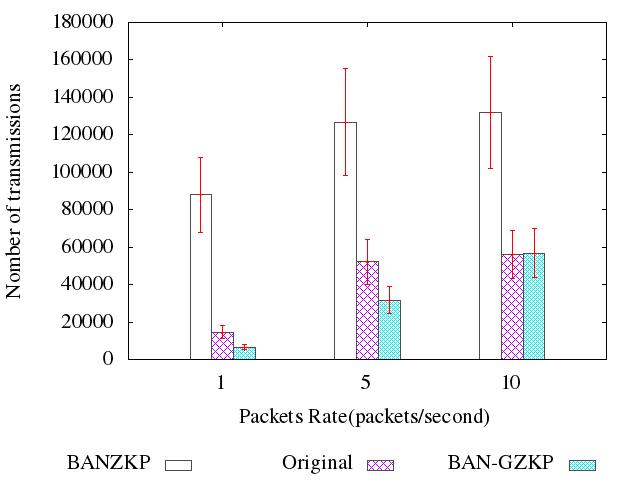}
\caption{Number of transmissions for FloodToSink in posture 3}
\label{weak}
\end{figure}

\begin{figure}
\centering
\includegraphics[width=0.4\textwidth]{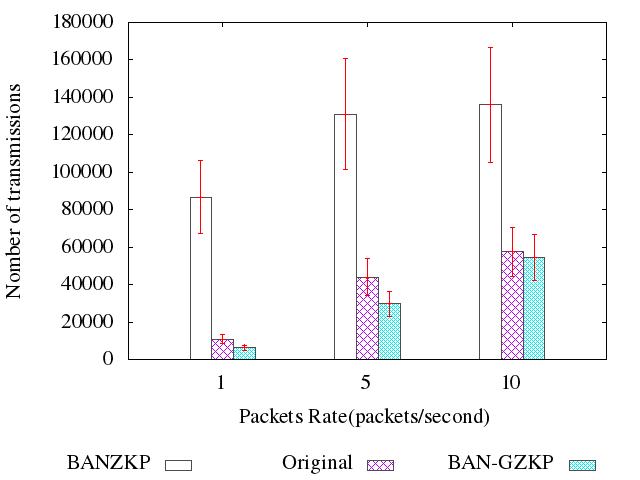}
\caption{Number of transmissions for FloodToSink in posture 4}
\label{sit}
\end{figure}

\begin{figure}
\centering
\includegraphics[width=0.4\textwidth]{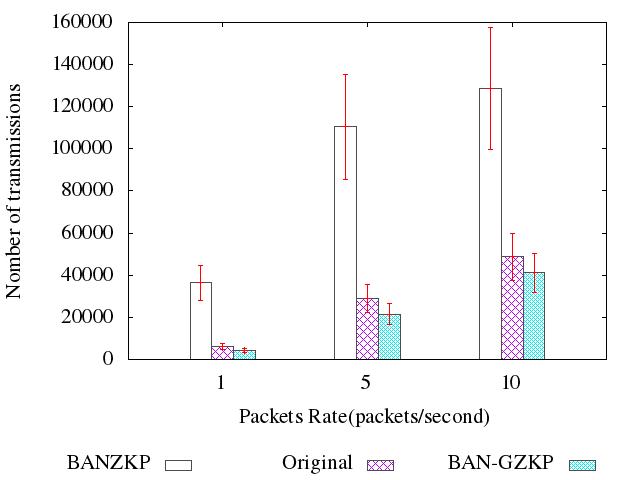}
\caption{Number of transmissions for FloodToSink in posture 5}
\label{lie}
\end{figure}

\begin{figure}
\centering
\includegraphics[width=0.4\textwidth]{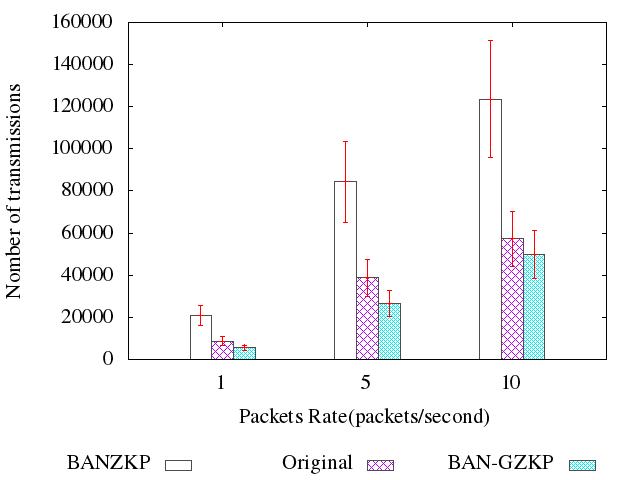}
\caption{Number of transmissions for FloodToSink in posture 6}
\label{sleep}
\end{figure}

\begin{figure}
\centering
\includegraphics[width=0.4\textwidth]{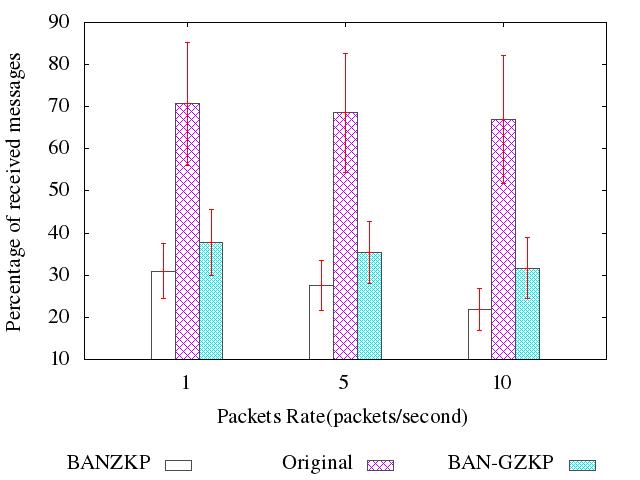}
\caption{Percentage of received messages for TreeBased in posture  1}
\label{Rp1TreeBased}
\end{figure}

\begin{figure}
\centering
\includegraphics[width=0.4\textwidth]{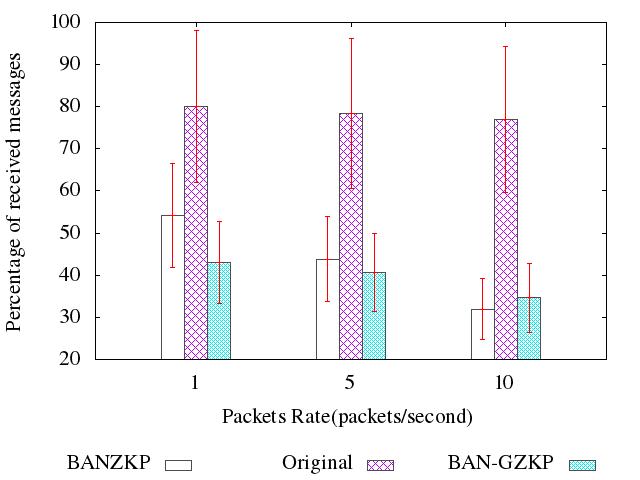}
\caption{Percentage of received messages for TreeBased in posture 3}
\label{Rp3TreeBased}
\end{figure}

\begin{figure}
\centering
\includegraphics[width=0.4\textwidth]{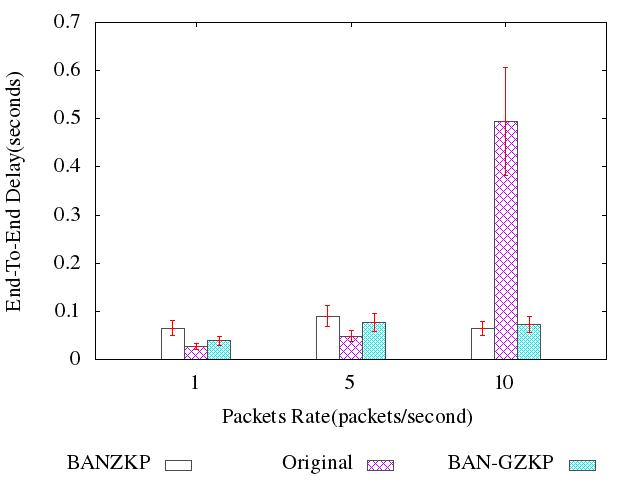}
\caption{End-To-End Delay for TreeBased in posture 6}
\label{Dp6TreeBased}
\end{figure}

\begin{figure}
\centering
\includegraphics[width=0.4\textwidth]{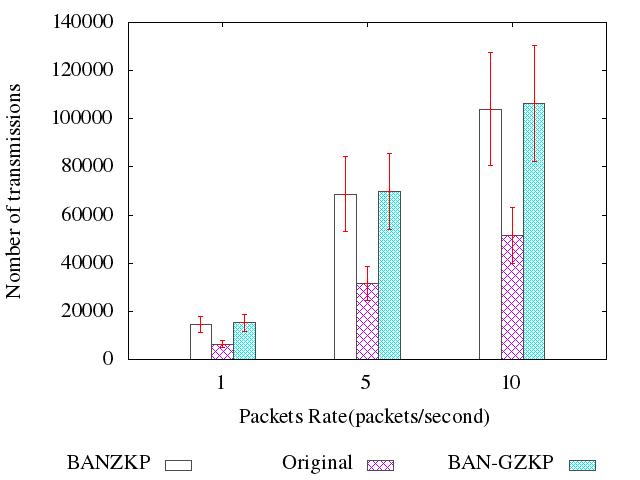}
\caption{Number of transmissions for TreeBased in posture 7}
\label{Tp7TreeBased}
\end{figure}

For \emph{TreeBased} strategy (see Figures from \ref{Rp1TreeBased} to \ref{Tp7TreeBased}, a example for posture 1 Walking), where packets will be re-sent several times according the links quality. The fast increase of end-to-end delay in case of original TreeBased strategy in postures 6 and 7 (see Figure \ref{Dp6TreeBased} for example). That because the general networks links quality are not good in posture 6 and 7 \cite{bu2017total}, the number of the retransmissions in these two postures are important, which lead to important collisions and backoffs with the increase of the packets generation rate. In general, BAN-GZKP has the lower end-to-end delay and number of transmissions in all the postures. In terms of ratio of packets reception, in posture 1, 2, 4 and 7 (see Figure \ref{Rp1TreeBased} for example), both BAN-GZKP and BANZKP are stable, and BAN-ZKP is better than BANZKP. However in postures 3 and 5 (see Figure \ref{Rp3TreeBased} for example), BANZKP is better when the packets generation rate is lower than 5 packets per second and 1 packet per second for postures 3 and 5, respectively ; when the packets generation rate is higher than 5 packets per second and 1 packet per second for postures 3 and 5, respectively, the performance of BANZKP decreases fast. And in posture 6, see Figure \ref{Rp6TreeBased}, both BANZKP and BAN-GZKP decrease fast when the packets generation rate is high. That is due to the retransmission mechanism of the TreeBased strategy, which lead to a high network burden. A flawed link not only lead to a decrease of the ratio of packets reception, but also an important number of retransmissions: that consumes the networks resource. If flawed links appear at the edge of the networks,  like in postures 1,2 and 7 (see \cite{bu2017total}), a hop-by-hop BAN-GZKP will limit the waste of the local retransmissions. But if the flawed links appear close to the sink, like in postures 3, 5 and 6, the BAN-GZKP could waste an important number of transmissions, according to the analyze  for strategy APAP and CTP. For the end to end BANZKP, the flawed links do not have an important influence. The defects will be enlarged by the retransmission mechanism of \emph{TreeBased} strategy. The performance of the BAN-GZKP has important variance in defects-enlarged links (by retransmission mechanism) if links defects occur close to the sink. 

\begin{figure}
\centering
\includegraphics[width=0.4\textwidth]{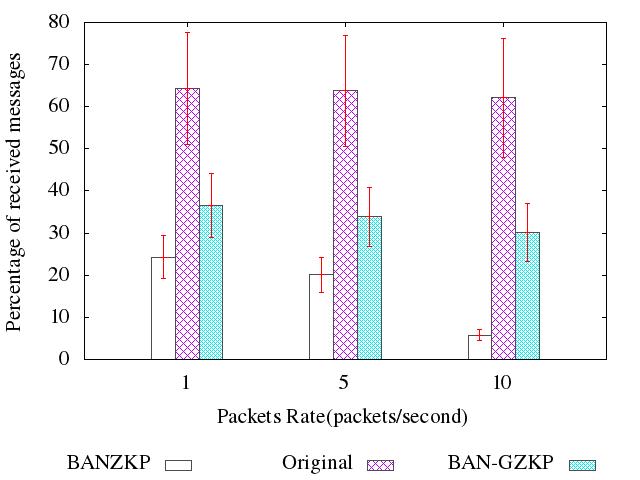}
\caption{Percentage of received messages for TreeBased in posture 2}
\label{run}
\end{figure}

\begin{figure}
\centering
\includegraphics[width=0.4\textwidth]{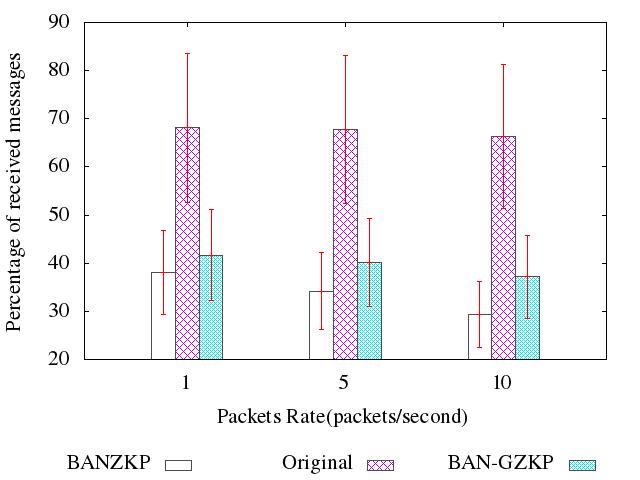}
\caption{Percentage of received messages for TreeBased in posture 4}
\label{sit}
\end{figure}

\begin{figure}
\centering
\includegraphics[width=0.4\textwidth]{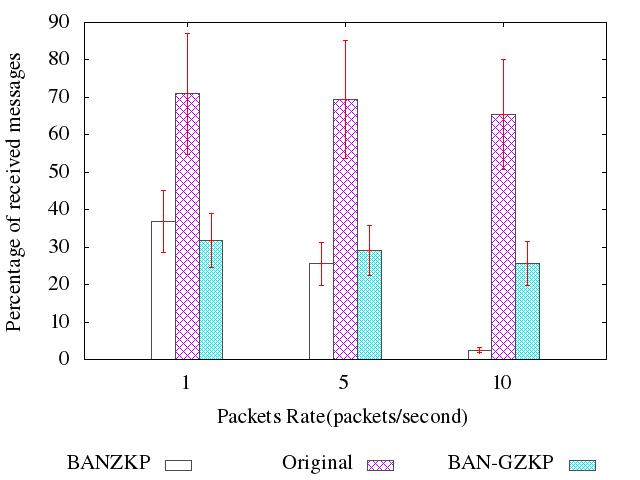}
\caption{Percentage of received messages for TreeBased in posture 5}
\label{lie}
\end{figure}

\begin{figure}
\centering
\includegraphics[width=0.4\textwidth]{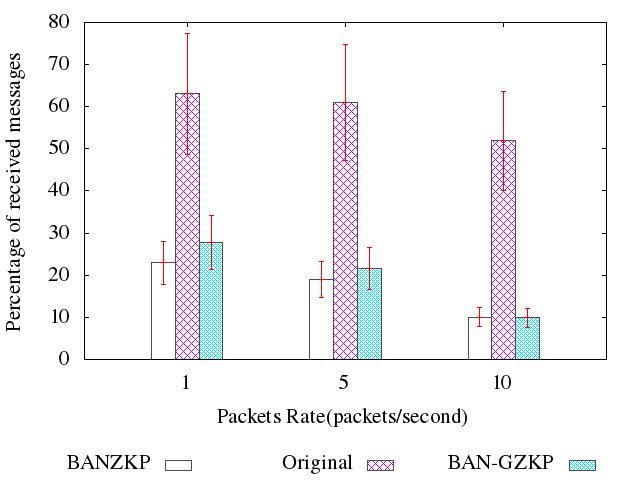}
\caption{Percentage of received messages for TreeBased in posture 7}
\label{jacket}
\end{figure}
\begin{figure}
\centering
\includegraphics[width=0.4\textwidth]{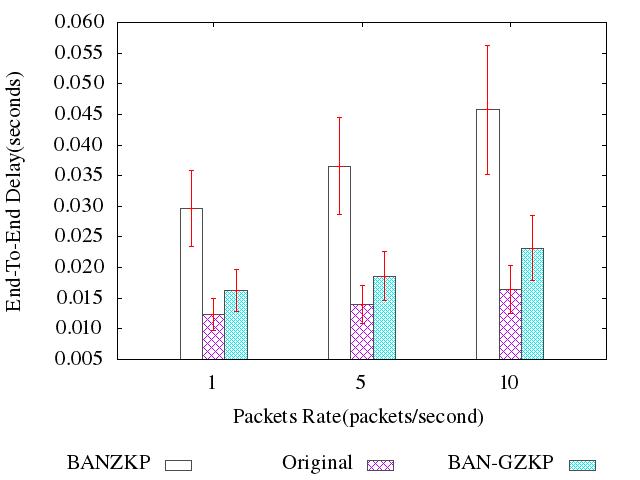}
\caption{End-To-End Delay for TreeBased in posture 1}
\label{walk}
\end{figure}

\begin{figure}
\centering
\includegraphics[width=0.4\textwidth]{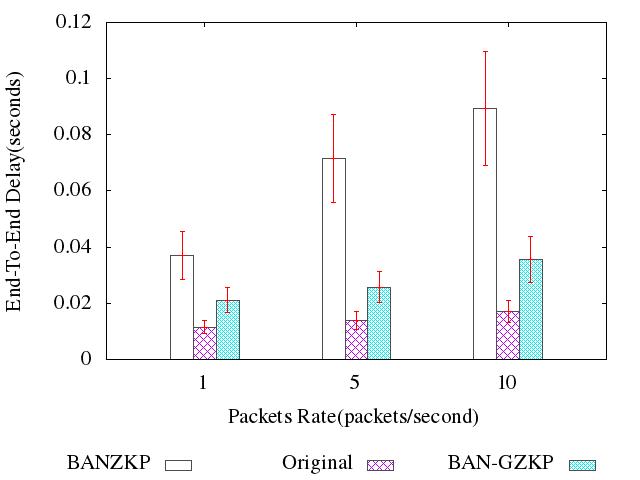}
\caption{End-To-End Delay for TreeBased in posture 2}
\label{run}
\end{figure}

\begin{figure}
\centering
\includegraphics[width=0.4\textwidth]{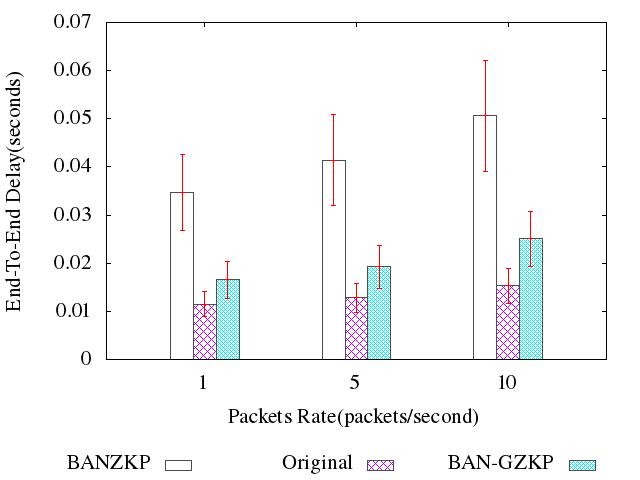}
\caption{End-To-End Delay for TreeBased in posture 3}
\label{weak}
\end{figure}

\begin{figure}
\centering
\includegraphics[width=0.4\textwidth]{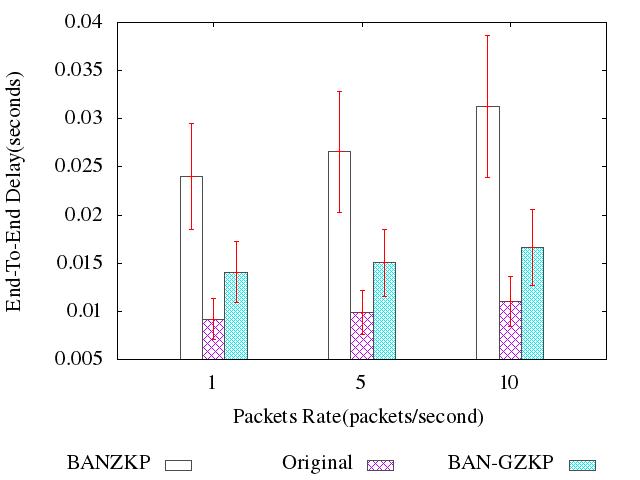}
\caption{End-To-End Delay for TreeBased in posture 4}
\label{sit}
\end{figure}

\begin{figure}
\centering
\includegraphics[width=0.4\textwidth]{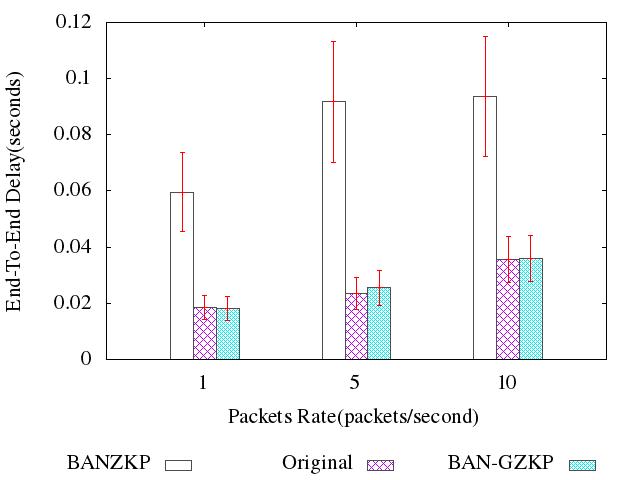}
\caption{End-To-End Delay for TreeBased in posture 5}
\label{lie}
\end{figure}

\begin{figure}
\centering
\includegraphics[width=0.4\textwidth]{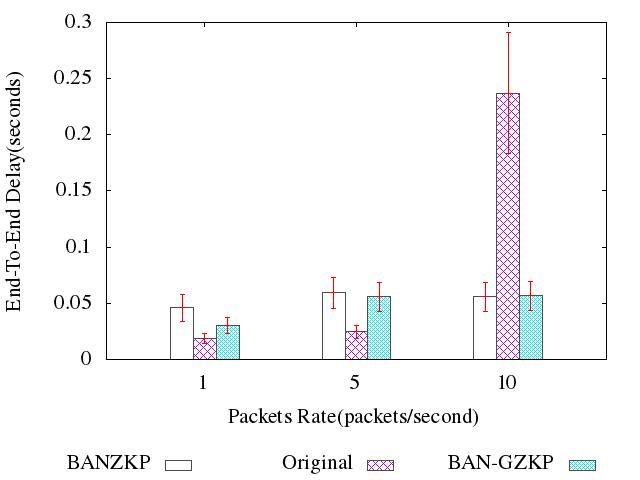}
\caption{End-To-End Delay for TreeBased in posture 7}
\label{jacket}
\end{figure}
\begin{figure}
\centering
\includegraphics[width=0.4\textwidth]{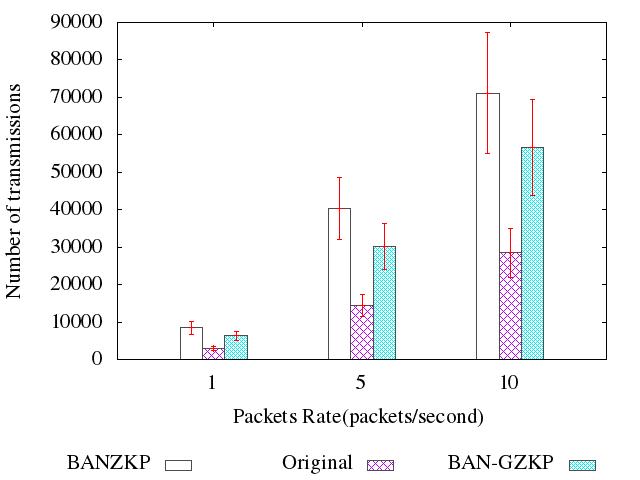}
\caption{Number of transmissions for TreeBased in posture 1}
\label{walk}
\end{figure}

\begin{figure}
\centering
\includegraphics[width=0.4\textwidth]{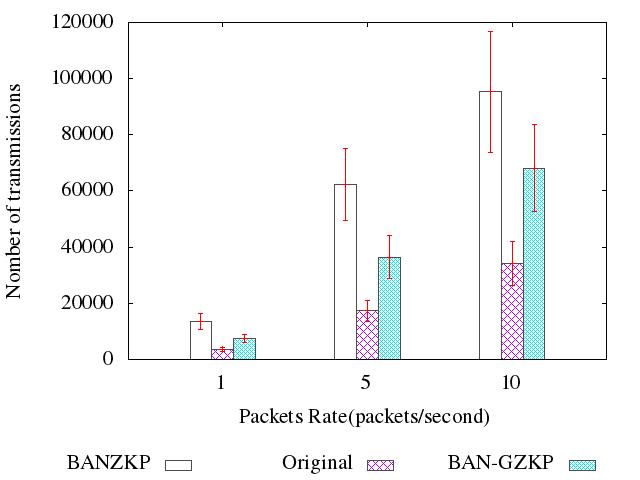}
\caption{Number of transmissions for TreeBased in posture 2}
\label{run}
\end{figure}

\begin{figure}
\centering
\includegraphics[width=0.4\textwidth]{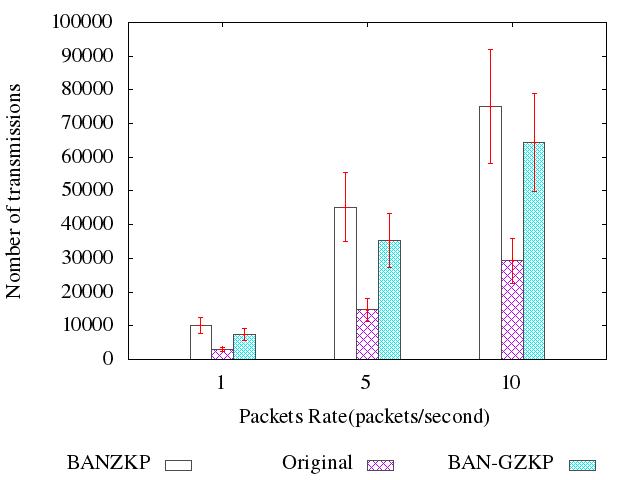}
\caption{Number of transmissions for TreeBased in posture 3}
\label{weak}
\end{figure}

\begin{figure}
\centering
\includegraphics[width=0.4\textwidth]{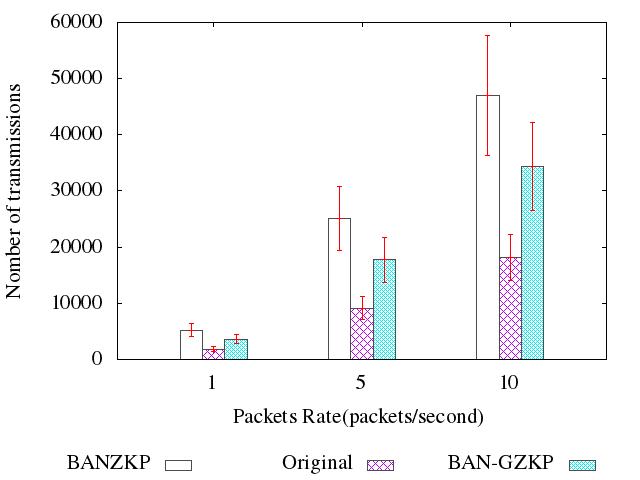}
\caption{Number of transmissions for TreeBased in posture 4}
\label{sit}
\end{figure}

\begin{figure}
\centering
\includegraphics[width=0.4\textwidth]{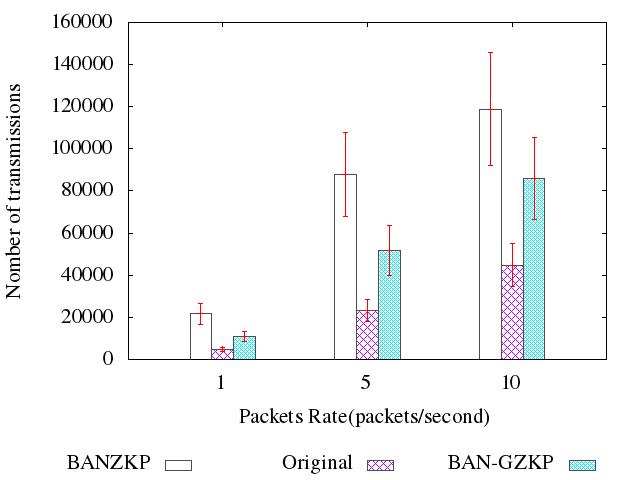}
\caption{Number of transmissions for TreeBased in posture 5}
\label{lie}
\end{figure}

\begin{figure}
\centering
\includegraphics[width=0.4\textwidth]{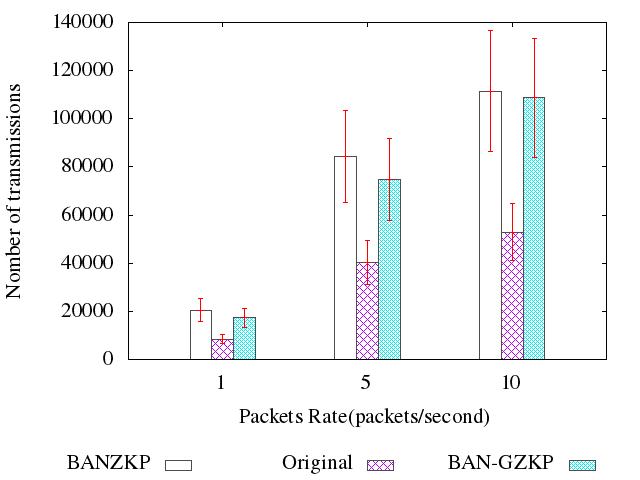}
\caption{Number of transmissions for TreeBased in posture 6}
\label{sleep}
\end{figure}

\section{Conclusion}
\label{sectionIV}
In this paper we proposed a new ZKP-based security scheme specifically designed for WBAN networks.  Our scheme, BAN-GZKP uses three ingredients: a novel \emph{random key allocation} which makes it resilient to the replay attack and redundancy information crack, a \emph{hop-by-hop} authentication scheme which makes it resilient to DDoS attacks at sink and a {ZKP exchanging optimisation} to further reduce the number of transmissions. Our BAN-GZKP improves, without any additional cost, the security level of the best ZKP scheme designed so far for WBAN networks. Moreover, when BAN-GZKP is used in order to secure existing convergecast protocols their performances are drastically improved compared to the case when BANZKP is used.





\bibliographystyle{elsarticle-num}

\bibliography{sample}

\end{document}